\documentclass[figuresleft]{aa}
\usepackage{natbib}
\usepackage{subcaption}

\usepackage{graphicx}
\usepackage{color}
\usepackage[table, dvipsnames]{xcolor}
\usepackage{hyperref}
\hypersetup{
    colorlinks=true,
    citecolor=blue,
    filecolor=black,
    linkcolor=blue,
    urlcolor=blue
}

\definecolor{highlight}{rgb}{1, 0.8, 0.8}

\usepackage[flushleft]{threeparttable}
\usepackage{multirow}
\usepackage{tablefootnote}
\usepackage{scrextend}
\usepackage{pdflscape}
\usepackage{graphicx}
\usepackage{rotating}

\usepackage{txfonts}
\usepackage[capitalise]{cleveref} 
\usepackage{placeins}

\defcitealias{2017A&A...604A..78R}{RT17}
\newcommand{\RT}{\citetalias{2017A&A...604A..78R}}

\begin{document}

   \title{Properties of intermediate- to high-mass stars in the young cluster M17}

   \subtitle{Characterizing the (pre-)zero-age main sequence}

   \author{F. Backs
          \inst{1, 2}
          \and
          S. A. Brands
          \inst{2}
          \and
          M. C. Ram\'irez-Tannus
          \inst{3}
          \and
          A. R. Derkink
          \inst{2}
          \and
          A. de Koter
          \inst{2}$^,$
          \inst{1}
          \and
          J. Poorta
          \inst{2}
          \and
          J. Puls
          \inst{4}
          \and
          Lex Kaper
          \inst{2}
          }

   \institute{Institute of Astronomy, KU Leuven, Celestijnenlaan 200D, 3001 Leuven, Belgium \\
              \email{frank.backs@kuleuven.be} \\
         \and
         Anton Pannekoek Institute for Astronomy, Universiteit van Amsterdam, Science Park 904, 1098 XH Amsterdam, The Netherlands
         \and
         Max-Planck Institut für Astronomie (MPIA), Königstuhl 17, D-69117 Heidelberg, Germany
         \and
         LMU M\"unchen, Universit\"atssternwarte, Scheinerstr. 1, D-81679 M\"unchen, Germany}

   \date{Received April 24, 2024; accepted August 6, 2024}

 
  \abstract
   {The outcome of the formation of massive stars is an important anchor point in the formation and evolution process of these stars. It provides insight into the physics of the assembly process, and sets the conditions for stellar evolution. For massive stars, the outcome of formation is rarely observed because the processes involved unfold deep down in highly extincted molecular clouds.} 
   {We characterize a population of highly reddened stars in the very young massive star-forming region M17. The group of 18 O4.5 to B9 stars constitutes one of the best samples of almost zero-age main-sequence (ZAMS) high- and intermediate-mass stars. Their properties allow us to identify the empirical location of the ZAMS of massive stars, and the rotation and properties of the mass-loss rate of stars close to or at the onset of core-hydrogen burning.}
   {We performed quantitative spectroscopic modeling of a uniform set of over 100 spectral features in optical VLT/X-shooter spectra using the nonlocal thermal equilibrium stellar atmosphere code {\sc Fastwind} and a fitting approach based on a genetic algorithm, {\sc Kiwi-GA}. The spectral energy distributions of photometric observations were used to determine the line-of-sight extinction. From a comparison of their positions in the Hertzsprung-Russell diagram with MIST evolutionary tracks, we inferred the stellar masses and ages. } 
   {We find an age of $0.4_{-0.2}^{+0.6}$\,Myr for our sample, however we also identify a strong relation between the age and the mass of the stars. All sources are highly reddened, with $A_V$ ranging from 3.6 to 10.6 mag. The sample can be subdivided into two groups. Stars more massive than 10\,M$_{\odot}$ have reached the ZAMS. Their projected ZAMS spin rate distribution extends to 0.3 of the critical velocity; their mass-loss rates agree with those of other main-sequence O and early-B stars. Stars with a mass in the range $3 < M/$M$_{\odot} < 7$ are still on the pre-main sequence (PMS), and some of them have circumstellar disks. Evolving their $\varv \sin i$ to the ZAMS assuming angular momentum conservation yields values up to $\sim 0.6 \,\varv_{\rm crit}$. For PMS stars without disks, we find tentative mass-loss rates up to $10^{-8.5}\,$M$_{\odot}$\,yr$^{-1}$. The total-to-selective extinction $R_V$ is higher for PMS stars with disks than for the remainder of the sample.}
   {We constrain the empirical location of the ZAMS for massive ($10 < M/$M$_{\odot} < 50$) stars and find it to agree with its location in MIST evolutionary tracks. The ZAMS rotation rates for intermediate-mass stars are twice as high as for massive stars, suggesting that the angular momentum gain processes differ between the two groups. The relation between the age and mass of the stars suggests a lag in the formation of more massive stars relative to lower mass stars.
   Taking the derived mass-loss rates at face value, stellar winds are already initiated in the PMS phase. The PMS-star winds are found to be substantially more powerful than indicated by predictions for line-driven outflows.}

   \keywords{Massive stars --
                Winds --
                Stellar atmospheres
               }

   \maketitle

\section{Introduction}

The past decades have shown tremendous progress in our understanding of low-mass star formation. The progress reaches from identifying morphological stages in the formation process and spectral accretion signatures to spatially resolved imaging of protoplanetary disks and a detailed modeling of both single and cluster star formation (see, e.g., Protostars and Planets VII \citeyear{2023ASPC..534.....I} for reviews of these subjects). This wealth of observations and models is not yet available for massive stars. The reasons are well known and include the short formation timescales involved and the assembly of massive stars happens in highly obscured molecular clouds. Moreover, the formation of massive stars is rare and therefore distant. Furthermore, the physical processes are complex due to radiative and mechanical feedback in the late stages of assembly \citep{2009Sci...323..754K,2010ApJ...722.1556K} and inevitable (close) binary or higher order multiple formation \citep[][]{2012Sci...337..444S,2022A&A...663A..26B}.

Fortunately, progress has also been impressive in our understanding of the formation of massive stars. Developments in the past decade seem to indicate that massive stars are formed through disk accretion similar to their lower-mass counterparts \citep[e.g][]{1987ARA&A..25...23S,2006Natur.444..703C,2009Sci...323..754K}. An increasing number of (interferometric) observations confirm the presence of accretion disks in several wavelength regimes \citep[e.g.][]{2016A&ARv..24....6B,2017NatPh..13..276C}. Furthermore, disk and companion formation is supported by the most advanced numerical simulations \citep[e.g][]{2018MNRAS.475.5618B,2020A&A...644A..41O,2023A&A...669A..80O}.

One route toward furthering our knowledge of the formation of massive stars is to identify the central sources of individual systems prior to or just at the arrival on the main sequence. As described above, this is extremely difficult because the sources are deeply embedded and emit the bulk of their radiation in the ultraviolet and optical wavelength domains, where extinction has the strongest impact. \citet{2020A&A...638A.157H} and \citet{2021A&A...646A.106S} emphasized the difficulty of detecting these stars as they find a significant dearth of massive stars near the zero-age main sequence (ZAMS) in their surveys. 
Characterizing the outcome of the formation of massive stars in terms of, for instance, the location of the ZAMS, rotational velocities, mass-loss rates (if winds have developed), multiplicity statistics, magnetic field strengths, and remnant disk properties (if any), is of great value for probing the formation mechanism. Moreover, these properties are important because they constitute the initial conditions of the subsequent evolution.

The youngest massive star-forming clusters are most suitable for this approach. One such young cluster is \object{NGC\,6618} located in \object{Messier 17} (\object{M17}). M17 is a well-studied giant \ion{H}{ii} region 
with a luminosity of $\sim 2.4\times10^6\,$L$_\odot$ \citep{2007ApJ...660..346P}. It is located at a distance of $1675^{+19}_{-18}$\,pc based on \emph{Gaia}-DR3 data \citep{2024A&A...681A..21S}. The \ion{H}{ii} region is surrounded by a giant molecular cloud with a gas content of approximately $6 \times 10^4\,$M$_\odot$ \citep{2009ApJ...696.1278P}. The cluster has a rich stellar population with $\sim$20 O-type stars and over 100 B-type stars \citep[][]{1997ApJ...489..698H,2008ApJ...686..310H,2024A&A...681A..21S}. A unique feature of M17 is that photospheric features have been detected along with hallmarks of circumstellar disks despite the severe extinction towards the sources \citep[][hereafter \RT]{2017A&A...604A..78R}. This has allowed the spectroscopic characterization of several of the sources and the confirmation of their pre-main-sequence (PMS) nature. The detected circumstellar disk features include CO-bandhead emission in several sources \citep[][also \RT]{1997ApJ...489..698H}. For some stars, this CO-bandhead emission has been modeled in detail, confirming the presence of hot dense gas close to the star \citep[][]{2023A&A...676A.122P}. This agrees with circumstellar hydrogen and dust-emission modeling that was undertaken for two sources \citep{2023A&A...671A..13B}. 

The presence of PMS stars suggests that the cluster is young. Age estimates yield $\sim$0.5\,Myr \citep{2008ApJ...686..310H} and $\sim$1\,Myr \citep{1997ApJ...489..698H,2009ApJ...696.1278P} based on the fraction of stars with IR-excess emission and isochrone fits. \citet{2024A&A...681A..21S} find an age of $0.65 \pm 0.25$\,Myr by tracing runaway stars back to their origin in NGC\,6618 with \emph{Gaia}\,DR3, and \citet{2014ApJ...787..108G} find an average age of $\sim$1.3\,Myr based on the X-ray luminosity of potential subclusters. For the latter study, the X-ray method yields individual subcluster ages in the range of 0.7 to 2.4\,Myr.

We study the stellar properties of a diverse population of 18 young high- and intermediate-mass stars in M17. The goals are to reassess the cluster age, to establish the empirical location of the ZAMS for massive stars, to constrain initial spin velocities, and, finally, to estimate the mass-loss properties, especially to establish whether PMS stars have already developed a stellar wind. When we limit the importance of these properties to star formation, the initial rotational velocity holds clues, for instance, for the angular momentum gain regulation by gravitational torques \citep{2011MNRAS.416..580L} or for the magnetic coupling between the star and the inner disk \citep{2010A&A...517A..58D}. The stellar wind, along with the photospheric radiation field, may ablate circumstellar disks \citep{2019MNRAS.483.4893K}, and on larger spatial scales, it plays a role in the self-regulation of massive star formation and the star formation efficiency of their host cloud \citep{2021ApJ...922L...3L,2022ApJ...941..202R}.

To reach our goals, we analyzed spectra obtained with the Very Large Telescope/X-Shooter of our target stars using stellar atmosphere models. The line-of-sight extinction is estimated by fitting spectral energy distributions (SED). This paper is set up as follows. First, we introduce the observations and data in Section\,\ref{p3:sec:data}. We describe the quantitative spectral analysis method in Section\,\ref{p3:sec:methods}. The results of the fitting are given in Section\,\ref{p3:sec:results} and are then discussed in Section\,\ref{p3:sec:discussion}. Our main conclusions are summarized in \cref{p3:sec:conclusion}.

\section{Targets and data}\label{p3:sec:data}

We briefly describe the sample and data. For a complete overview of the spectroscopic data, we refer to \citet{M17_binaries}

\subsection{Sample}

Table\,\ref{p3:tab:sample} lists the sources we analyze in this paper. The sample was selected such that it contained some of the most massive stars in M17 and a population of PMS stars \citep{1997ApJ...489..698H,2013ApJS..209...32B,2017A&A...604A..78R}. The sample was selected to be diverse, and it is not complete. The spectral types range from O4.5 to B9. Together with the strongly varying extinction toward the sources, this results in a sample spanning over seven $V$-band magnitudes.
Table\,\ref{p3:tab:sample} also indicates the detected excess emission for the stars with circumstellar disks. Some stars show thermal infrared (IR) continuum emission by dust and gas line emission, others only show IR excess emission. For most stars, no excess emission was detected.

\begin{table}
\centering
\caption{List of the analyzed M17 sources.}
\label{p3:tab:sample}
\def\arraystretch{1.1}
\tiny
\begin{tabular}{llcccc} \hline \hline
Source & SpT & RA & DEC & $V$ [mag] & Emission \\ \hline
B111   & O4.5 V     & 18:20:34.65 & -16:10:11.3 & 11.3 & -- \\
B164   & O6 Vz      & 18:20:31.03 & -16:10:08.9 & 16.9 & -- \\
B311   & O8.5 Vz    & 18:20:22.90 & -16:08:34.3 & 13.7 & -- \\
B181   & O9.7 III   & 18:20:30.39 & -16:10:36.2 & 17.4 & -- \\
B289   & O9.7 V     & 18:20:24.43 & -16:08:45.8 & 17.0 & IR \\
B215   & B1 V       & 18:20:28.55 & -16:12:13.5 &  --  & IR \\
B93    & B2 V       & 18:20:35.80 & -16:10:54.7 &  --  & -- \\
B205   & B2 V       & 18:20:29.11 & -16:11:13.3 & 16.1 & -- \\
CEN55  & B2 III     & 18:20:29.08 & -16:10:57.2 & 18.0 & -- \\
B234   & B3 IV      & 18:20:27.42 & -16:10:11.6 & 18.0 & -- \\
B213   & B3 IV      & 18:20:28.61 & -16:09:24.2 & 17.7 & -- \\
B253   & B4 III     & 18:20:26.26 & -16:11:04.4 & 15.7 & -- \\
B150   & B5 V       & 18:20:31.78 & -16:11:40.5 & 15.6 & IR$^{\dagger}$ \\
B272   & B7 III     & 18:20:25.30 & -16:09:42.7 & 18.2 & -- \\
B275   & B7 III     & 18:20:25.19 & -16:10:25.4 & 15.6 & Gas and IR \\
B243   & B9 III     & 18:20:26.41 & -16:10:04.8 & 17.8 & Gas and IR \\
B268   & B9 III     & 18:20:25.25 & -16:10:21.6 & 17.1 & Gas and IR \\
B86    & B9 III     & 18:20:36.16 & -16:10:57.4 & 14.6 & -- \\
\hline
\end{tabular}
\tablefoot{
    The final column indicates the circumstellar excess emission as identified in \RT. IR indicates that continuum IR excess emission is detected, and gas indicates that gas phase emission lines are detected in the optical and NIR spectra. \\
    $^\dagger$ Result from this work.}
\end{table}

\subsection{Spectra}

The data consist of multi-epoch X-shooter spectra obtained from May to September 2019 with additional archival data from 2012 and 2013. We only use the UVB and VIS spectral arms of X-shooter here, which were set to slit widths of 1" and 0.9", respectively. This yields a resolving power of 5100 from $\sim$3500 to 5500\,\AA\ and 8800 from 5500 to 10\,000\,\AA.

The data were reduced using the ESO pipeline \texttt{esorex~3.3.5} \citep{2010SPIE.7737E..28M}. Relative flux calibrated data were obtained using spectrophotometric standards from the ESO database, and telluric corrections were made with \texttt{Molecfit~1.5.9} \citep{2015A&A...576A..78K}. In the case of B243, B268, and B275, a special reduction was done to correct for severe nebular contamination \citep[see][]{2020A&A...636A..54V,2024A&A...681A.112D}.

The signal-to-noise ratio of the spectra varies strongly between stars because their brightnesses are different, but it also varies as a function of wavelength as a result of the strong extinction toward the stars. Therefore, the signal-to-noise ratio can be 10--20 at the shortest wavelengths, but may reach >100 in the VIS arm.

\subsection{Photometry}

Accurate photometric data are required in order to constrain the luminosity and line-of-sight extinction toward the sources
(see Section\,\ref{p3:sec:extinction}). We adopted the photometry presented in \RT. For the stars that are not covered in \RT, we used the photometric data from {\sc simbad}\footnote{http://simbad.cds.unistra.fr/simbad/}, with UVB data from \citet{2003AJ....125.2531R}, JHK data from 2MASS \citep{2006AJ....131.1163S}, and mid-infrared data from \emph{Spitzer} IRAC \citep{2003PASP..115..953B}. For all stars, a \emph{Gaia} G-band point was also added \citep{2021A&A...649A...3R}.

\section{Methods} \label{p3:sec:methods}

We aimed to determine accurate stellar atmosphere properties of the sample of (pre-)main-sequence sources in M17. To do this, we fit {\sc Fastwind} \citep{1997A&A...323..488S,2005A&A...435..669P} stellar atmosphere models to the normalized stacked spectra. 
The genetic algorithm {\sc Kiwi-GA}\footnote{https://github.com/sarahbrands/Kiwi-GA} \citep{2022A&A...663A..36B} was used as the optimization method to find the best-fitting parameters and their uncertainties. 
Section~\ref{p3:sec:data_stack} describes the stacking process of the multi-epoch spectra. The extinction determination is described in Section~\ref{p3:sec:extinction} and the normalization process is explained in Section~\ref{p3:sec:normalization}. Finally, {\sc Fastwind} and {\sc Kiwi-GA} are described in Sections~\ref{p3:sec:fastwind} and \ref{p3:sec:GA}, respectively.

\subsection{Data stacking}\label{p3:sec:data_stack}

The multi-epoch spectra presented in \citet{M17_binaries} were stacked to improve the signal-to-noise ratio. We stacked the relative flux-calibrated spectra to preserve the uncertainties on the data and to minimize potential biases from manipulating the data. First, we used the radial velocities determined by \citet{M17_binaries} to shift the observed spectra to the same rest frame. Then the spectra were placed in the same wavelength binning using linear interpolation. Since the observing conditions can vary between the different observation epochs, the continuum levels of the relative flux calibrated spectra did not always match. This can be problematic when a weighted mean for the stacked spectrum is to be determined. Therefore, the continuum levels were matched first. This was done by dividing one observed spectrum by the other. This resulted in a featureless residual that represents the difference between the continuum levels. The residual was then fit using a linear function. In some cases, an issue occurred in the flux calibration, resulting in a nonlinear residual. When this occurred, a seventh-order polynomial was fit instead. The relative flux calibrated spectrum was multiplied by the fit to the residual so that it matched the other observed spectra. This procedure was repeated for each observed spectrum, such that the continuum levels of all epochs matched. The flux-matched spectra were then stacked using weights defined as  
\begin{equation}
    w_{\lambda, i} = \frac{\sigma_{\lambda, i}^{-1}}{\sum_i^n \sigma_{\lambda, i}^{-1}},
\end{equation}
with $\sigma_{\lambda, i}$ the uncertainty on the flux at wavelength (bin) $\lambda$ and epoch $i$, and $n$ the number of observing epochs. The weighted mean flux was determined as 
\begin{equation}
    \bar{\mathcal{F}_{\lambda}} = \sum_i^n w_{\lambda, i}~\mathcal{F}_{\lambda, i},
\end{equation}
and the corresponding uncertainties as
\begin{equation}
    \sigma_{\lambda} = \sum_i^n w_{\lambda, i}^2~\sigma_{\lambda, i}^2.
\end{equation}
This resulted in a higher signal-to-noise ratio in the relative flux-calibrated spectrum compared to the individual single-epoch spectra. The stacked spectrum was then dereddened as the extinction properties are essential in the determination of the stellar luminosity (see \cref{p3:sec:extinction}). This also aided in the normalization process (see Section~\ref{p3:sec:normalization}) and in the potential identification of IR-excess emission.

\subsection{Determining extinction and luminosity}\label{p3:sec:extinction}

Accurate knowledge of the line-of-sight extinction is required to determine the stellar luminosity. Additionally, it aids in the normalization process because the shape of the spectrum is more predictable. The extinction was determined by fitting a stellar model SED multiplied by an extinction curve to the photometric SED. The stellar model SEDs we used are from \citet{2004A&A...419..725C} for the approximate temperature of the star based on their spectral type\footnote{The applied conversion of spectral type into temperature is from \url{http://www.isthe.com/chongo/tech/astro/HR-temp-mass-table-byhrclass.html}}. The model SEDs were convolved with the filter profile of the respective photometric filters. For stars with near-IR excess, only photometric data points with wavelengths shorter than the onset of the excess emission were considered (see \cref{p3:sec:result_extinction}).

We used the extinction curve of \citet{1999PASP..111...63F}, with the total-to-selective extinction, $R_V$, and the total visual extinction, $A_V$, as free parameters. $R_V$ was fit because the stars are located in a dense star-forming region, in which its value may be higher than the canonical value of $R_V = 3.1$ \citep[see e.g.][]{1975A&A....43..133S}. In the SED fitting, we also used a third free parameter, $c$, which represents the approximate radius of the star when it is scaled with the distance to M17. The fitting was done in a two-step approach. First, the best-fit parameters were determined with an optimization routine in the {\sc scipy} package for {\sc python}, then a 2D brute-force grid search in $R_V$ and $A_V$ was applied around the best-fit parameters to verify their values, uncertainties, and correlation.

The best-fitting extinction curve was then used to determine the absolute J-band magnitude, which served as the luminosity anchor: The SED calculated by {\sc Fastwind} was compared to this luminosity anchor to determine the stellar radius. Longer wavelengths are less strongly affected by extinction. We chose to use the $J$ band as it is the longest wavelength filter that is not affected by the IR-excess emission from circumstellar material. 

\subsection{Normalization and data-clipping process}\label{p3:sec:normalization}

The observed relative flux-calibrated stacked spectra were normalized before the fitting process. Normalization was made after stacking to minimize the biases and irregularities caused by the normalization process, and it was applied locally around the modeled features, which are listed in Section\,\ref{p3:sec:GA}. A linear function was fit to continuum points selected by eye, and the spectrum was divided by this linear fit.

As a result of the high extinction and because the stars reside in an \ion{H}{ii} region, the spectra contain many interstellar features. These were removed to prevent them from affecting the fitting process. The clipped features include diffuse interstellar bands (DIBs), nebular hydrogen and helium lines, disk emission lines, stellar lines that are not included in the model, and instrumental artifacts. The clipping was done manually by inspecting the spectra. DIBs were mostly identified by their similar behavior in all spectra, but also by consulting the list of DIBs from \citet{2009ApJ...705...32H}.

\subsection{Fastwind}\label{p3:sec:fastwind}

We determined the stellar photosphere and wind properties simultaneously, maintaining a uniform fitting approach for the full sample. Our stellar model of choice was the nonlocal thermal equilibrium (non-LTE) stellar atmosphere code {\sc Fastwind} \citep{1997A&A...323..488S,2005A&A...435..669P,2018A&A...619A..59S}, version 10.6. {\sc Fastwind} treats both the stellar atmosphere and the trans-sonic outflow. The atomic model is split into explicit and background elements, the combination of which is used to treat the impact of a multitude of lines on radiation transport and back-heating (so-called line blocking and line blanketing). Only the radiative transfer for the explicit elements is treated in the comoving frame. In our {\sc Fastwind} version, these explicit elements are H, He, C, N, O, and Si, and thus, we can produce synthetic line profiles only for these elements. Not all ionization stages were included for the explicit ions (to reduce computation cost). For this reason, some of the cooler stars have prominent lines that cannot be modeled. The formal solution of the radiative transfer equation is limited to specific requested lines or sets of lines. These sets or blends of lines we refer to as "complexes".

\subsection{Fitting approach and genetic algorithm}\label{p3:sec:GA}

\begin{table}
\centering
\caption{Diagnostic features used in the fitting process.}
\small
\label{p3:tab:diagnostics}
\begin{tabular}{llc} \hline \hline
Ion & Wavelength [\AA] & Part of complex \\  \hline
\ion{Si}{ii}    & 3853.7, 3856.0, 3862.6  & \multirow{2}{*}{H$\zeta$}           \\
\ion{H}{i}      & 3889.1                  &            \\ \hline
\ion{O}{iii}    & 3961.6                  & \multirow{3}{*}{H$\epsilon$}            \\
\ion{He}{i}     & 3964.7                  &             \\
\ion{H}{i}      & 3970.1                  &             \\ \hline
\ion{He}{ii}    & 4025.7                  & \multirow{2}{*}{\ion{He}{i}\,4026}         \\
\ion{He}{i}     & 4026.3                  &           \\ \hline
\ion{C}{iii}    & 4068.9, 4070.3          & \multirow{5}{*}{\ion{C}{iii}\,4070}        \\
\ion{O}{ii}     & 4069.6, 4069.9, 4072.2  &         \\
                & 4075.9                  &                 \\
\ion{O}{iii}    & 4072.6, 4074.0          &         \\
\ion{H}{i}      & 4101.8                  &         \\ \hline
\ion{Si}{iv}    & 4088.9                  & \multirow{7}{*}{H$\delta$}          \\
\ion{N}{iii}    & 4097.4                  &           \\
\ion{He}{ii}    & 4100.1                  &           \\ 
\ion{H}{i}      & 4101.8                  &           \\
\ion{N}{iii}    & 4103.4                  &           \\
\ion{Si}{iv}    & 4116.1                  &           \\
\ion{Si}{ii}    & 4128.1, 4130.9          &           \\ \hline
\ion{He}{i}     & 4143.8                  & \ion{He}{i}\,4143         \\ \hline
\ion{C}{iii}    & 4186.9                  & \multirow{4}{*}{\ion{He}{ii}\,4200}        \\
\ion{N}{iii}    & 4195.8, 4200.1, 4215.8  &         \\
\ion{He}{ii}    & 4199.9                  &         \\
\ion{Si}{iv}    & 4212.4, 4212.4          &         \\ \hline
\ion{C}{ii}     & 4267.0, 4267.3, 4267.3  & \ion{C}{ii}\,4267         \\ \hline
\ion{O}{ii}     & 4317.1, 4319.6, 4366.9  & \multirow{4}{*}{H$\gamma$}          \\
\ion{N}{iii}    & 4332.9, 4337.0, 4345.7  &           \\
\ion{He}{ii}    & 4338.8                  &           \\
\ion{H}{i}      & 4340.5                  &           \\ \hline
\ion{N}{iii}    & 4379.0, 4379.2          & \multirow{2}{*}{\ion{He}{i}\,4387}         \\
\ion{He}{i}     & 4388.0                  &          \\ \hline
\ion{He}{i}     & 4471.5                  & \ion{He}{i}\,4471         \\ \hline
\ion{N}{iii}    & 4510.9, 4511.0, 4514.9  & \multirow{2}{*}{\ion{N}{iii}\,4515}        \\
                & 4518.1, 4523.6          &                 \\ \hline
\ion{N}{iii}    & 4534.6                  & \multirow{3}{*}{\ion{He}{ii}\,4541}        \\
\ion{He}{ii}    & 4541.7                  &         \\
\ion{Si}{iii}   & 4552.6                  &         \\ \hline
\ion{Si}{iii}   & 4567.8, 4574.8          & \ion{Si}{iii}\,4570       \\ \hline
\ion{N}{iii}    & 4634.1, 4640.6, 4641.9  & \ion{O}{ii}\,\ion{N}{iii}\,46      \\
\ion{O}{ii}     & 4638.9, 4641.8          & \ion{O}{ii}\,\ion{N}{iii}\,46      \\ \hline
\ion{C}{iii}    & 4647.4, 4650.2, 4651.5  & \ion{C}{iii}\,\ion{Si}{iv}\,46      \\
\ion{Si}{iv}    & 4654.3                  & \ion{C}{iii}\,\ion{Si}{iv}\,46      \\ \hline
\ion{O}{ii}     & 4661.6, 4676.2          & \ion{O}{ii}\,\ion{C}{iii}\,46      \\
\ion{C}{iii}    & 4663.6, 4665.9          & \ion{O}{ii}\,\ion{C}{iii}\,46      \\ \hline
\ion{He}{ii}    & 4685.9                  & \ion{He}{ii}\,4686        \\ \hline
\ion{He}{i}     & 4713.2                  & \ion{He}{i}\,4713         \\ \hline
\ion{N}{iii}    & 4858.7, 4859.0, 4861.3  & \multirow{5}{*}{H$\beta$}           \\
                & 4867.1, 4867.2, 4873.6  &                 \\
                & 4884.1                  &                 \\
\ion{He}{ii}    & 4859.4                  &            \\
\ion{H}{i}      & 4861.4                  &             \\ \hline
\ion{He}{i}     & 4922.0                  & \ion{He}{i}\,4922         \\ \hline
\ion{He}{i}     & 5015.7                  & \ion{He}{i}\,5015         \\ \hline
\ion{Si}{ii}    & 5041.0, 5056.0, 5056.3  & \multirow{2}{*}{\ion{He}{i}\,5047}         \\
\ion{He}{i}     & 5047.8                  &          \\ \hline
\ion{He}{ii}    & 5411.6                  & \ion{He}{ii}\,5411        \\ \hline
\ion{O}{iii}    & 5592.4                  & \ion{O}{iii}\,5592        \\ \hline
\ion{He}{i}     & 5875.7                  & \ion{He}{i}\,5875         \\ \hline
\ion{Si}{ii}    & 6347.1, 6371.4          & \ion{Si}{ii}\,6347        \\ \hline
\ion{He}{ii}    & 6560.2                  & \multirow{3}{*}{H$\alpha$}          \\
\ion{H}{i}      & 6562.8                  &           \\
\ion{C}{ii}     & 6578.1, 6582.9          &           \\ \hline
\ion{He}{i}     & 6678.2                  & \multirow{2}{*}{\ion{He}{ii}\,6683}        \\
\ion{He}{ii}    & 6683.3                  &         \\ \hline
\ion{Si}{iv}    & 7047.9, 7047.9, 7068.4  & \multirow{2}{*}{\ion{He}{i}\,7065}         \\
\ion{He}{i}     & 7065.3                  &          \\ \hline
\end{tabular}
\end{table}

The stellar properties we studied are the effective temperature of the star, $T_{\rm eff}$; the surface gravity, $g$; the mass-loss rate, $\dot{M}$; the projected surface rotational velocity, $\varv \sin i$; the helium surface abundance, $y_{\rm He}$, and the surface abundance of C, N, O, and Si. This results in a total of nine free parameters. The microturbulent velocity was held fixed at 10\,km\,s$^{-1}$. This value may be on the high side for late B-type stars, which might affect the abundance determinations \citep[e.g.][]{2022ApJ...937..110L}. We did not include a macroturbulent velocity field, which accounts for motions in the photosphere other than thermal motion, microturbulent velocities, and rotation. The flow velocity profile is given by a standard $\beta$-type velocity law, with $\beta = 1$. Optical diagnostic features have limited sensitivity to the terminal velocity $\varv_{\infty}$ of the outflow, and therefore, it cannot be constrained. We used measurements of \citet{2023arXiv230312165H} and the spectral type to estimate the $\varv_{\infty}$ values. 
We assumed that optically thin clumps may be present in the outflow, characterized by a clumping factor $f_{\rm cl} = 10$.

The nine free parameters were to be fit simultaneously due to several correlations between the parameters. The exploration of this parameter space can be computationally expensive. Therefore, the optimization of parameter values was done using {\sc Kiwi-GA} \citep{2022A&A...663A..36B}, building on the work of \citet{2005A&A...441..711M,2011ApJ...741L...8T}. This is a genetic algorithm (GA) and wrapper tailored to {\sc Fastwind}. The GA starts with randomly and uniformly sampling a specified parameter space. This random sampling is the first generation of models, after which the next generations are determined based on the fitness of the models in the generations before them. This works by combining the parameters of two of the best-fitting models up to that moment, and the better-fitting models have a higher probability to reproduce. The parameter values are mixed. Some values come from the one parent model, and others from the other model. Parameters can also mutate, in which case their value does not match either of the parents' values. Two types of mutation are included, which we call "broad" and "narrow". The broad mutation has a low probability and samples the new parameter value from a broad Gaussian distribution spanning a significant part of the parameter space. The narrow mutation has a higher probability and is sampled from a narrow Gaussian distribution. Both distributions are centered around the value inherited from the parent. These mutations ensure a proper exploration of the parameter space. Details of all the meta-parameters regarding the fitting algorithm can be found in \citet{2022A&A...663A..36B}.

{\sc Kiwi-GA} is designed to fit a set of normalized spectral diagnostic features. We opted to fit the same set of features for all stars, regardless of their temperature or spectral type. This was done to ensure that differences between stars are not the result of the line selection. The features were selected based on their strength, lack of contamination from unknown lines and features, and established reliability as a diagnostic in the temperature range probed by our stars. The full list of diagnostic features is given in Table~\ref{p3:tab:diagnostics}. Because the line selection is uniform for all stars, some fit lines are absent from the observed spectrum. 
As the selected lines are free of blends with lines that are not modeled, this should not be a problem.  
For example, the selected \ion{He}{ii} complexes show only noise or modeled features (of other species) in the cooler (late-B) stars. Therefore, it should not impact the fitting process adversely. 

The uncertainties on the model parameters are based on their fitness, which was assessed using either a $\chi^2$ statistic or a root-mean-square-error-of-approximation statistic \citep[RMSEA;][]{RMSEA_PAPER}. If the reduced $\chi^2$ of the best-fit model was below 1.5, the former was used, and otherwise, the RMSEA was used as it is designed to produce reasonable uncertainties if the models do not perfectly reproduce the observations. More details of the RMSEA uncertainty determination can be found in \citet{Brands2023_inprep}. Regardless of the method, the uncertainties should be considered estimates of the 1$\sigma$ confidence intervals. 

\section{Results} \label{p3:sec:results}

We present the main results of our fitting efforts below. \cref{p3:sec:result_extinction} shows our findings regarding the line-of-sight extinction properties. The atmosphere and wind analysis results are described in \cref{p3:sec:result_atmosphere}, and some parameters are highlighted. Poorer fits are mentioned in \cref{p3:sec:result_anomalous_fits}.

\subsection{Extinction}\label{p3:sec:result_extinction}

The results of the SED fitting are listed in Table~\ref{p3:tab:extinction_results} along with the approximate onset wavelength of the IR excess. The IR excess of six stars mandates the exclusion of data points at wavelengths beyond the IR cutoff specified in \cref{p3:tab:extinction_results} from the  extinction fits. The visual line-of-sight extinction to the sources varies from 3.6 to 10.6 magnitudes, and the total-to-selective extinction varies from 2.7 to 5.7. Only for B213 and B215 did we find an $R_V$ value below 3. This could be due to the limited photometric data available. Additionally, the $V$-band photometry for B213 appears to be significantly brighter than the rest of the photometric data. Both stars, and particularly B215, show a strong correlation between $A_V$ and $R_V$ (see Appendix~\ref{p3:sec:app_extinction}). Full SEDs along with the dereddened SEDs and correlations between $A_V$ and $R_V$ are provided in Appendix~\ref{p3:sec:app_extinction}.

\begin{table}
\centering
\caption{Extinction properties and luminosity anchor, $J_{\rm abs}$. }
\small
\label{p3:tab:extinction_results}
\def\arraystretch{1.2}
\begin{tabular}{lcccc} \hline \hline
Source & IR cutoff & $A_V$ & $R_V$ & $J_{\rm abs}$\\ \hline
B111 & ... & $5.19\pm0.02$ & $3.52\pm0.02$ & $-4.34\pm0.07$\\
B164 & ... & $8.37\pm0.02$ & $4.05\pm0.03$ & $-3.29\pm0.07$\\
B311 & ... & $6.23\pm0.03$ & $3.65\pm0.04$ & $-2.98\pm0.07$\\
B181 & ... & $10.55\pm0.06$ & $4.65\pm0.17$ & $-3.29\pm0.07$\\
B289 & 3.0 $\mu$m & $8.13\pm0.03$ & $4.14\pm0.04$ & $-2.80\pm0.07$\\
B215 & 3.0 $\mu$m & $9.14\pm0.30$ & $2.71\pm0.22$ & $-2.30\pm0.11$\\
B93  & ... & $3.70\pm0.03$ & $4.00\pm0.20$ & $-0.92\pm0.07$\\
B205 & ... & $7.51\pm0.04$ & $3.68\pm0.06$ & $-1.71\pm0.07$\\
CEN55& ... & $9.24\pm0.05$ & $3.80\pm0.06$ & $-1.60\pm0.08$\\
B234 & ... & $8.40\pm0.04$ & $3.72\pm0.06$ & $-0.83\pm0.07$\\
B213 & ... & $8.51\pm0.05$ & $2.73\pm0.04$ & $-1.46\pm0.07$\\
B253 & ... & $6.73\pm0.04$ & $3.90\pm0.08$ & $-1.46\pm0.07$\\
B150 & 3.0 $\mu$m & $5.40\pm0.04$ & $4.29\pm0.14$ & $-0.41\pm0.07$\\
B272 & ... & $9.04\pm0.04$ & $3.06\pm0.03$ & $-1.24\pm0.07$\\
B275 & 1.5 $\mu$m & $7.41\pm0.07$ & $4.69\pm0.11$ & $-2.59\pm0.07$\\
B243 & 1.5 $\mu$m & $7.92\pm0.06$ & $5.68\pm0.21$ & $-0.94\pm0.07$\\
B268 & 1.5 $\mu$m & $7.51\pm0.07$ & $4.95\pm0.19$ & $-1.29\pm0.08$\\
B86  & ... & $3.61\pm0.02$ & $3.58\pm0.14$ & $0.16\pm0.06$\\
\hline
\end{tabular}
\tablefoot{IR cutoff indicates the approximate wavelength of the onset of IR excess emission, if present. }
\end{table}

\subsection{Atmosphere fits} \label{p3:sec:result_atmosphere}

\begin{figure*}
    \centering
    \includegraphics[width=0.85\textwidth]{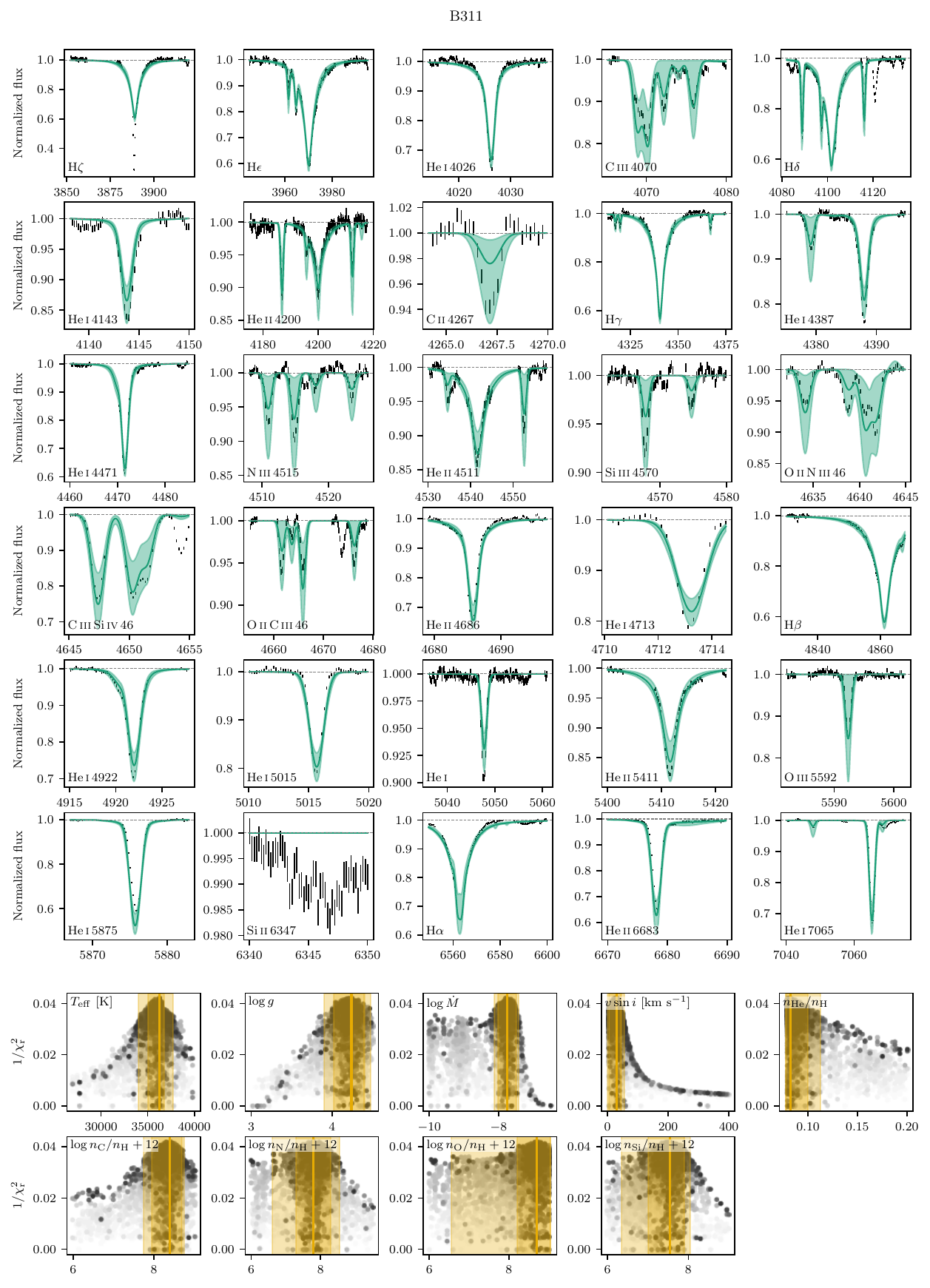}
    \caption{Overview of the fit to B311. The \emph{top} part shows the fits to each diagnostic. The vertical black bars indicate the normalized observed spectra, and the green line shows the best-fit model. The length of the black bars indicate the 1$\sigma$ uncertainty on the data, and the shaded region shows the 1$\sigma$ confidence interval of the model spectrum. The name of the diagnostic feature is given at the bottom of each panel. The \emph{bottom} part shows the parameter distributions sampled by {\sc Kiwi-GA}. The fit parameters are displayed in the top left corner. The shade of the points is a measure of the generation: darker points are sampled in later generations. The vertical yellow line shows the best-fit value, the darker yellow shaded region shows the 1$\sigma$ confidence interval, and the lighter yellow shaded region shows the 2$\sigma$ confidence interval. }
    \label{p3:fig:fit_summary_B311}
\end{figure*}

\cref{p3:fig:fit_summary_B311} show the best model fits to the diagnostic line profiles and their uncertainties for B311. The other overviews of the other stars are in \cref{p3:sec:fit_summaries} and available online\footnote{Available at \url{https://doi.org/10.5281/zenodo.13285505}}. The confidence intervals of the fit parameters are shown in the bottom two rows. A condensed overview of best-fit line profiles and their uncertainties is shown in \cref{p3:app:results}. 
The corresponding stellar parameters are listed in Table~\ref{p3:tab:fit_results}. The sample consists of stars with a wide range of temperatures and signal-to-noise ratios. The best-fit temperatures span from $\sim$11\,kK to $\sim$46\,kK and the luminosities from $10^{2}$\,L$_\odot$ to $3\times10^5$\,L$_\odot$. Below, we discuss the main findings of the fitting efforts per parameter.

\subsubsection{Temperature and gravity}

The surface temperature and gravity are mostly determined by the hydrogen and helium lines and by the line-strength ratios between them, with additional constraints from the C, N, O, and Si lines. Throughout the sample, the temperature is well constrained, with larger relative uncertainties at lower temperatures. This is likely due to a lower number of diagnostic lines available and to the lower signal-to-noise ratio of the fainter stars. Gravity determinations are robust, with well-constrained values in the range $\log g = 3.60$ -- 4.24, which result in spectroscopic masses that are consistent with their evolutionary masses (see also \cref{p3:app:mass_discrepancy}). The gravity determination of B272 is an outlier in the sample, with $\log g = 4.86$, which is significantly higher than for the other stars. This is likely due to the poor quality of the spectrum of this star.

\subsubsection{Mass-loss rates}

The mass-loss rate determination is mostly based on the wind emission in H$\alpha$ and to a lesser extent in H$\beta$; \ion{He}{ii}\,4686 adds a mass-loss constraint for the O-type stars. For the hotter more luminous stars, the mass-loss rate determinations are robust and reliable. However, at a lower luminosity ($L \lesssim 10^4\,$L$_\odot$), the winds are weaker and therefore harder to constrain. We present the mass-loss rates and their uncertainties as produced by our fitting routine and the corresponding statistics. We stress that the inferred mass-loss rates may be strongly affected by systematic uncertainties. The values should be treated with care as the sensitivity to this parameter is limited at these weak wind strengths. Best-fit values can change significantly through small inconsistencies in the data. We note that we did not consider mass-loss rates $\dot{M} < 10^{-10}$\,M$_\odot$\,yr$^{-1}$ in our modeling and assumed a wind clumping factor of $f_{\rm cl} = 10$.

\subsubsection{Rotation rates}

The projected rotation rate, $\varv \sin i$, of the stars is typically well constrained. Cooler stars and stars with a low signal-to-noise ratio have a less accurate $\varv \sin i$. This is particularly true for stars with a gaseous circumstellar disk or significant nebular contamination, as in these cases, the central part of the strongest lines had to be clipped from the data. 
Two stars, CEN55 and B272, show a best-fit $\varv \sin i$ of 5\,km\,s$^{-1}$, which is the lowest value we considered. These lines might be affected by nebular lines that are blended with the stellar line profiles (see \cref{p3:sec:anom_fit_CEN55} and \cref{p3:sec:anom_fit_B272}). Additionally, the low-quality spectrum of B272 might affect the derived rotation rate.

\subsubsection{Abundances} \label{p3:sec:results_abundance}

The helium abundances of the stars are typically well determined, with increasing uncertainties toward lower temperatures because the helium features are less pronounced. However, it has been challenging to constrain the surface abundance of C, N, O, and Si because only a few strong diagnostic features are available in our uniform modeling approach, the signal-to-noise ratio of some of the spectra is low, and the rotation rate of many of the stars is high. Combined with the correlation and degeneracy of the abundances with some of the other atmosphere parameters, this resulted in large uncertainties for most stars. The exception to this is B311, which has a high signal-to-noise ratio and a low projected rotation rate. This star has well-constrained abundances, the best-fit values of which agree very closely with the values presented in \citet{2009ARA&A..47..481A}. This is further discussed in Section~\ref{p3:sec:abundance_discussion}.

\subsection{Anomalous fits} \label{p3:sec:result_anomalous_fits}

Some best-fit solutions appear to be anomalous and should be interpreted with caution. We list these cases here.

\subsubsection{CEN55}\label{p3:sec:anom_fit_CEN55}

We find a good fit for this star for a very low projected rotation rate, $\varv \sin i$. However, because of the narrow stellar line profiles, the nebular contributions to the \ion{He}{i} lines are hard to distinguish from the stellar lines. Therefore, the fit might be contaminated, resulting in an odd helium surface abundance. The highest helium abundance considered in our parameter space is the best fit, but it is not expected to be enhanced as the stars in M17 have just formed.

\subsubsection{B272}\label{p3:sec:anom_fit_B272}

B272 has the highest $V$-band magnitude in the sample. The spectra have a poor signal-to-noise ratio throughout most of our wavelength range. Similar to CEN55, this star has a best-fit value of $\varv\sin\,i=5$\,km\,s$^{-1}$ and a significantly enhanced helium abundance. We also find a high surface gravity of $\log g = 4.68$, which is significantly higher than expected for a star of this luminosity and temperature. The spectroscopic mass corresponding to this is 34.4\,M$_\odot$, which is also higher than expected (see \cref{p3:app:mass_discrepancy}). 

\subsubsection{B181}

B181 shows stronger \ion{He}{ii} lines than were found in the best-fit model. This may suggest that the temperature of the star is underestimated. However, temperatures that match the \ion{He}{ii} line strengths fall within the confidence interval. A higher temperature would place this star closer to the ZAMS. 

\section{Discussion} \label{p3:sec:discussion}

The young age of M17 allowed us to investigate near-ZAMS properties of a sample of stars covering masses from about 2--30\,M$_{\odot}$. For the most massive stars, we can identify the empirical location of the ZAMS, and for the full sample, we can place constraints on (pre-)ZAMS rotational velocities and mass-loss rates. 
We first assess the evolutionary status of the sample (\cref{p3:sec:HRD}) to establish stellar ages and masses, and we highlight a correlation between these two properties (\cref{p3:sec:age_determination}). Next, we discuss their present-day and (future) ZAMS surface rotation (\cref{p3:sec:rotation}). We compare our results to those of \RT\ in \cref{p3:sec:RT_comparison}. Then, we briefly discuss the surface abundances (\cref{p3:sec:abundance_discussion}) and line-of-sight extinction (\cref{p3:sec:extinction_discussion}). Finally, we discuss the mass-loss properties (\cref{p3:sec:mass_loss_rates}). We summarize our results in \cref{p3:sec:conclusion}.

\subsection{Evolutionary status of the sample} \label{p3:sec:HRD}

\begin{figure}
    \centering
    \includegraphics[width=\columnwidth]{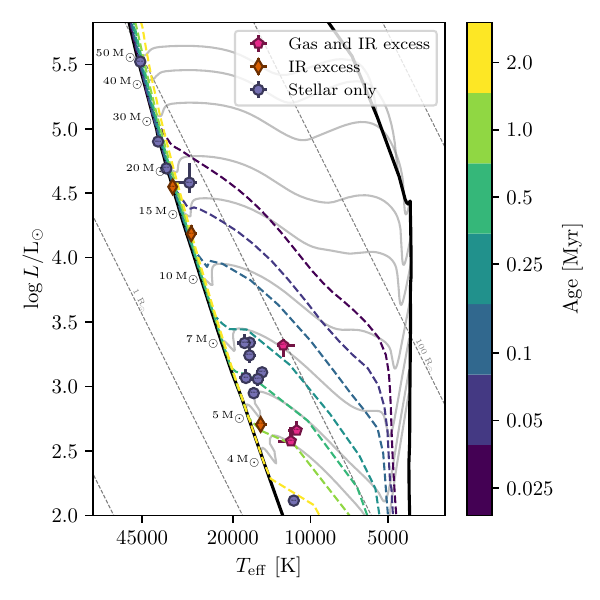}
    \caption{Hertzsprung-Russell diagram of the sample of M17 stars. The stars are separated into three categories, depending on their circumstellar emission features. Pink pentagons stars show both emission from a gaseous disk and IR excess emission from hot circumstellar dust. Orange indicates stars that only show IR excess, and purple indicates that no circumstellar material was detected here or by \RT. The MIST PMS evolutionary tracks are overplotted in gray, and the masses are labeled on the left. We show the ZAMS in black (on the left) and the birthline (on the right) \citep{2016ApJS..222....8D,2016ApJ...823..102C}. The isochrones (with t=0 at the birthline) are shown with the dashed colored lines. The thin dashed gray  lines show the isoradius lines to guide the eye. The names of the stars are displayed on the points and are visible when zooming in. }
    \label{p3:fig:HRD}
\end{figure}

Figure\,\ref{p3:fig:HRD} shows the locations of the stars in the Hertzsprung-Russell diagram (HRD). The MESA Isochrones and Stellar Tracks (MIST) PMS evolutionary tracks \citep{2016ApJS..222....8D,2016ApJ...823..102C} based on MESA models \citep{2011ApJS..192....3P} are overplotted for masses ranging between 3 and 50\,M$_\odot$.  We first examine the population of stars that are more massive than 13\,M$_{\odot}$. As we show below, they are all severely extincted, with $A_{V}$ ranging from 3.6 to 10.6\,mag. This would essentially render them invisible at optical wavelengths if not for the use of an 8m class telescope. Five out of six stars agree very well with the predicted location of the ZAMS, and one star (B181) is positioned slightly to the red. We might be finding massive stars (20--50\,M$_\odot$) on the ZAMS in contrast to \citet{2020A&A...638A.157H} and \citet{2021A&A...646A.106S} because we considered highly extincted objects. As the PMS evolution proceeds fast for these sources (it takes them $\sim 10^{5}$\,yr at most to reach the ZAMS) and the moment of core-hydrogen ignition signifies the transition from the blueward to redward evolution, this empirically confirms the location of the ZAMS in this mass range. In the tracks at hand, the ZAMS is reached when the hydrogen-burning luminosity exceeds 99.99\% of the total luminosity \citep{2016ApJS..222....8D}.

Our sample does not contain stars with masses in the range 7 -- 13\,M$_{\odot}$. 
While the sample was not actively selected to avoid this mass range, it is conceivable that stars of these masses are lacking as a result of our selection criteria. In selecting targets, we aimed to include the brightest sources and also PMS sources. Stars in the mass range 7 -- 13\,M$_\odot$ were likely not bright enough or not red enough to be selected. A uniform sampling may have been further hindered because the extinction varies for each line of sight. 

Stars with masses between about 3 and 7\,M$_{\odot}$ have not yet reached the main sequence. This is in line with previous studies, which indicated that the stars are in the final phase of their formation \citep[e.g.][]{2017A&A...604A..78R}. Stars with double-peaked line emission disks are located farthest from the main sequence, suggesting that these are the least evolved objects in the sample and have not yet lost their accretion disk. The three stars with IR excess emission are located close to or on the main sequence, indicating that in some cases, circumstellar dust can persist until the end of the formation process. However, this is not ubiquitous because most stars, including less evolved stars, do not show any excess emission.

\subsection{Age determination}\label{p3:sec:age_determination}

The isochrones in Fig.~\ref{p3:fig:HRD} already give an indication of the cluster age. However, a more accurate estimate may be obtained from determining the ages of the individual stars. To do this, we interpolated the evolutionary tracks and determined the range of ages that fall within the uncertainties on $L$ and $T_{\rm eff}$ of the stars. Through the same method, the evolutionary masses of the stars were determined. Table\,\ref{p3:tab:extra_parameters} lists the results. For stars that have reached the main sequence, the age uncertainties will be relatively large as evolution proceeds on a nuclear timescale. For stars on the PMS, the characteristic timescale at which they move along the tracks is the much shorter Kelvin-Helmholtz timescale. Therefore, we excluded the main-sequence sources from the cluster age determination. This results in a cluster age of $0.4_{-0.2}^{+0.6}$\,Myr.
This result matches the dynamical age of $0.65 \pm 0.25$\,Myr derived from tracing O- and early B-type runaway stars back to NGC\,6618 \citep{2024A&A...681A..21S} very well and also matches the approximate estimates by \citet{1997ApJ...489..698H,2008ApJ...686..310H,2009ApJ...696.1278P}.

The ages we derived based on PMS evolution are measured from the birthline, which in the MIST tracks is defined to be the locus at which the central stellar temperature reaches $10^5$\,K. At the birthline, the stars have accreted their final mass. 
We remark that although ages in PMS tracks are usually measured starting from the birthline, the definition of the birthline is somewhat ambivalent. It is often associated with the star becoming visible at optical light, linking it with the end of the main accretion phase, that is, when the dusty natal cloud from which the star obtains its mass is depleted or dispersed. Computations applying this alternative definition reported that massive stars reach the ZAMS while still accreting material \citep{2009ApJ...691..823H,2010ApJ...721..478H}. \citet{1990ApJ...360L..47P} involved the size of the proto-star in the definition of the birthline. It is therefore important to realize that estimated age may vary as a result of different initial conditions and definitions of the birthline \citep{2011ApJ...738..140H,2017A&A...599A..49K}.

\begin{figure}
    \centering
    \includegraphics[width=\columnwidth]{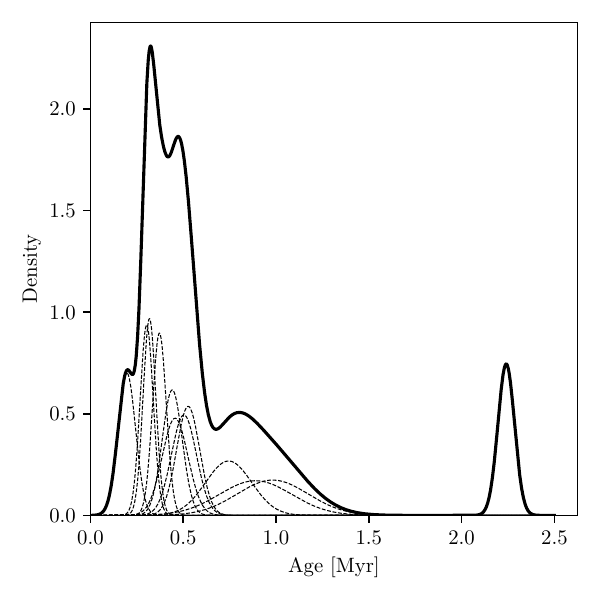}
    \caption{Distribution of PMS ages of the stars in M17 since they passed the birthline. The solid black line indicates an approximate kernel density estimate based on the best-fit ages and their confidence intervals. The thin dashed lines indicate the contribution of the individual sources. Higher-mass stars ($M>10$\,M$_\odot$) that have reached the main sequence are excluded. }
    \label{p3:fig:age_dist}
\end{figure}

\begin{figure}
    \centering
    \includegraphics[width=\columnwidth]{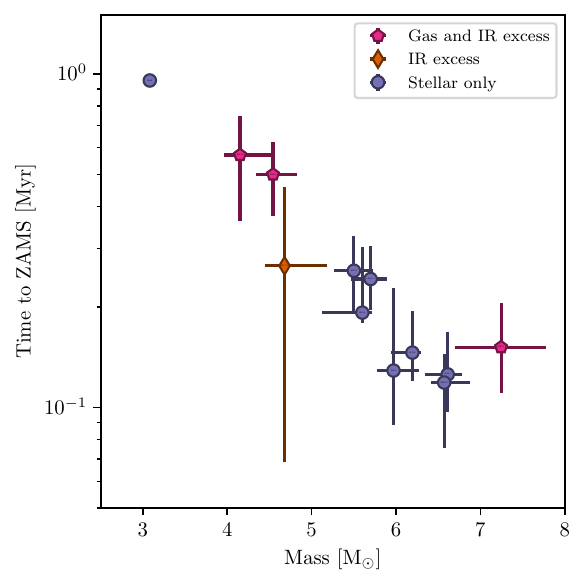}
    \caption{Time remaining for stars to reach the ZAMS as function of mass for the PMS stars in M17. The markers are identical to those in \cref{p3:fig:HRD}.}
    \label{p3:fig:time_to_ZAMS}
\end{figure}

The MIST tracks allow us to determine the time remaining for the stars to reach the ZAMS. This timescale is independent of the definition of the birthline and mainly depends on further Kelvin-Helmholtz contraction. It is given in \cref{p3:fig:time_to_ZAMS} and reveals that the most massive stars are closest in time to reaching the main sequence. Lower-mass stars have longer to go. 

\begin{figure}
    \centering
    \includegraphics[width=\columnwidth]{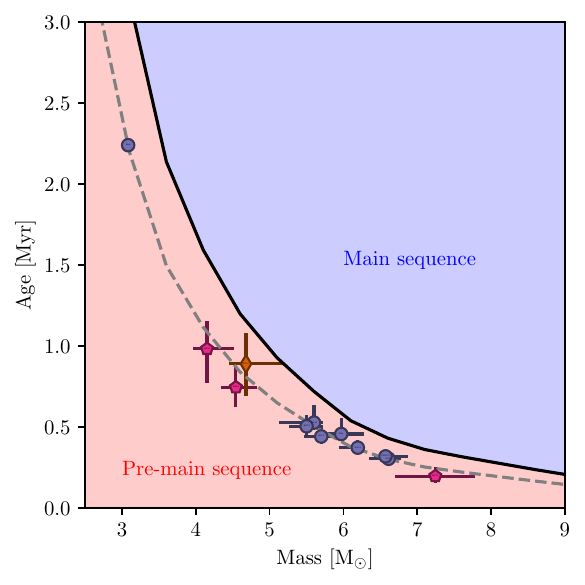}
    \caption{PMS age as function of mass measured from the moment the stars
    pass the birthline. The black line indicates the total time it takes to reach the ZAMS, starting from the birthline, and the  dashed gray line shows 70\% of that time. The markers are the same as in Figure\,\ref{p3:fig:HRD}.}
    \label{p3:fig:mass_vs_age}
\end{figure}

However, we not only find a correlation between the time remaining to reach the ZAMS and stellar mass, but also between age and stellar mass. This is shown in \cref{p3:fig:mass_vs_age}, which shows a trend for the PMS sources of a decreasing PMS age for increasing mass. This trend is consistent with the older ages found for the low-mass stars by \citet{2014ApJ...787..108G}. 
All stars appear to have completed $\sim$70\% of the time needed to reach the ZAMS, which is remarkable. This implies that in absolute time, low-mass stars cross the birthline first.
A similar trend is found when using PARSEC \citep{2012MNRAS.427..127B} PMS evolutionary tracks instead of MIST tracks. 

In the context of this finding, we recall that the accretion timescale of a higher-mass star might be longer than that of a lower-mass star. The considered PMS evolutionary tracks assume that all mass is accreted before the birthline is crossed. The time required to collect this mass from the start of the collapse of the cloud is not included. This time might explain the relation between age and mass shown in \cref{p3:fig:mass_vs_age}.

\begin{figure}
    \centering
    \includegraphics[width=\columnwidth]{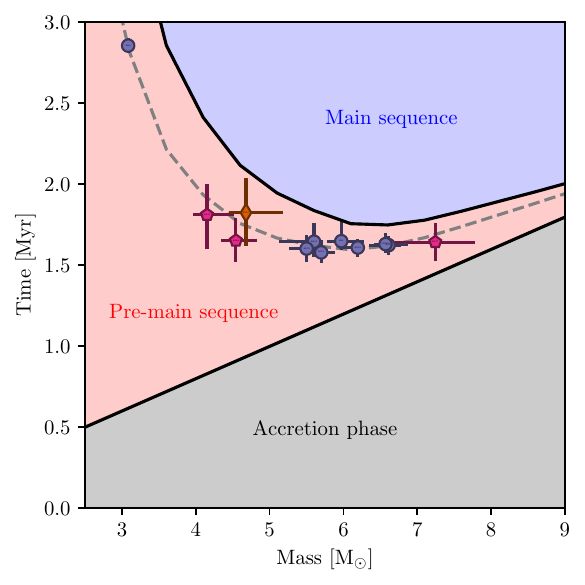}
    \caption{Time since the beginning of the main accretion phase for different stellar masses. The gray  area marks the main accretion phase with a constant accretion rate for all masses. The red area shows the PMS phase in which the star contracts to the main sequence. Blue indicates the main sequence. The dashed gray  line indicates 70\% of the PMS phase, as in \cref{p3:fig:mass_vs_age}. The data points indicate the time it would have taken the stars to first accrete their mass and then contract to their current state. }
    \label{p3:fig:mass_age_with_accretion_phase}
\end{figure}

The formation process of massive stars is complex. We considered a simplified scenario here in which stars accrete their mass and cross the birthline after they reach their final mass. \citet{1977ApJ...214..488S} showed that the mass-accretion rate scales with the (third power of the) sound speed in the collapsing natal cloud. A constant cloud temperature then implies the same mass-accretion rate for all forming stars.
When we assume an average accretion rate of $\dot{M} \sim 5 \times 10^{-6}$~M$_\odot$, we obtain the accretion timescale as shown by the bottom black line in \cref{p3:fig:mass_age_with_accretion_phase}. After this accretion phase, the stars contract to the main sequence through their PMS evolutionary tracks. The time this takes is given by the red shaded region. For the adopted accretion rate, all stars (except for B86, which appears to be older) would have started to form, that is, started to accrete material, at the same time $\sim$1.5\,Myr ago.
At this accretion rate, it would not be possible to form the highest-mass stars. For example, it would take nearly 10 Myr to form the most massive star in our sample, B111. This suggests that for the highest-mass objects, the accretion mechanics may be different, for instance, due to higher local gas cloud temperatures.

Regarding the star-forming history, we mention the possibility that multiple populations might be present in M17. \citet{2014ApJ...787..108G} identified different subclusters within M17 that, using an X-ray luminosity based age determination method, may span $\sim$2 Myr in age. An age spread does not directly explain the mass -- age relation, but it signals that the sample might be more complex. Furthermore, \citet{2016ApJ...819..139B} and \citet{2024A&A...682A.180D} reported high degrees of fragmentation in the dark cloud G14.225–0.506, which is located in the M17 complex. This suggests that new star formation might be taking place, and stars might only start to accrete matter now. Possibly in favor of a longer accretion and formation timescale for higher mass stars, \citet{2016ApJ...825..125P} reported a dearth of O-type stars in this dark cloud, given the large observed population of intermediate-mass PMS stars. They noted that the cloud contains enough mass to produce massive star clusters. Forming high-mass stars might still be deeply embedded in the cloud.%

\begin{table*}
\centering
\caption{Derived properties of stars in M17 based on evolutionary tracks. }
\label{p3:tab:extra_parameters}
\small
\def\arraystretch{1.5}
\begin{tabular}{lccccc} \hline \hline
Source & Age [Myr]& $M_{\rm ZAMS} [$M$_\odot]$ & $R_{\rm ZAMS} [$R$_\odot]$ & $\varv\sin i / \varv_{\rm crit}$ & $\varv_{\rm ZAMS}\sin i / \varv_{\rm crit}$ \\ \hline
B111     & $0.01^\uparrow_{...}$ & $47.4^{+1.9}_{-3.0}$ & $8.82^{+0.20}_{-0.30}$ & $0.15^{+0.02}_{-0.01}$ & $0.15^{+0.02}_{-0.01}$\\
B164     & $0.03^\uparrow_{...}$ & $25.7^{+1.1}_{-1.1}$ & $6.49^{+0.11}_{-0.16}$ & $0.10^{+0.01}_{-0.01}$ & $0.10^{+0.01}_{-0.01}$\\
B311     & $0.04^\uparrow_{...}$ & $20.3^{+0.6}_{-0.6}$ & $5.67^{+0.11}_{-0.08}$ & $0.03^{+0.01}_{-0.02}$ & $0.03^{+0.01}_{-0.02}$\\
B181     & $0.04^\uparrow_{...}$ & $16.9^{+3.3}_{-1.1}$ & $5.17^{+0.48}_{-0.18}$ & $0.19^{+0.03}_{-0.03}$ & $0.28^{+0.06}_{-0.06}$\\
B289     & $0.04^\uparrow_{...}$ & $17.9^{+0.8}_{-0.5}$ & $5.33^{+0.13}_{-0.08}$ & $0.16^{+0.02}_{-0.02}$ & $0.16^{+0.02}_{-0.02}$\\
B215     & $0.06^\uparrow_{...}$ & $13.3^{+1.0}_{-0.9}$ & $4.53^{+0.18}_{-0.15}$ & $0.24^{+0.02}_{-0.03}$ & $0.24^{+0.02}_{-0.03}$\\
B93      & $0.46^{+0.10}_{-0.04}$ & $6.0^{+0.3}_{-0.2}$ & $2.89^{+0.09}_{-0.06}$ & $0.33^{+0.03}_{-0.03}$ & $0.41^{+0.05}_{-0.06}$\\
B205     & $0.30^{+0.04}_{-0.03}$ & $6.6^{+0.2}_{-0.3}$ & $3.08^{+0.04}_{-0.08}$ & $0.29^{+0.04}_{-0.04}$ & $0.50^{+0.10}_{-0.09}$\\
CEN55    & $0.32^{+0.03}_{-0.04}$ & $6.6^{+0.3}_{-0.2}$ & $3.07^{+0.07}_{-0.05}$ & $0.01^{+0.03}_{-0.01}$ & $0.01^{+0.05}_{-0.01}$\\
B234     & $0.53^{+0.11}_{-0.01}$ & $5.6^{+0.1}_{-0.5}$ & $2.78^{+0.03}_{-0.14}$ & $0.46^{+0.05}_{-0.03}$ & $0.60^{+0.09}_{-0.07}$\\
B213     & $0.37^{+0.05}_{-0.03}$ & $6.2^{+0.1}_{-0.2}$ & $2.96^{+0.03}_{-0.08}$ & $0.35^{+0.03}_{-0.03}$ & $0.55^{+0.07}_{-0.06}$\\
B253     & $0.44^{+0.06}_{-0.05}$ & $5.7^{+0.2}_{-0.2}$ & $2.81^{+0.06}_{-0.07}$ & $0.26^{+0.09}_{-0.06}$ & $0.47^{+0.19}_{-0.12}$\\
B150     & $0.89^{+0.19}_{-0.20}$ & $4.7^{+0.5}_{-0.2}$ & $2.51^{+0.15}_{-0.07}$ & $0.41^{+0.03}_{-0.04}$ & $0.51^{+0.07}_{-0.07}$\\
B272     & $0.50^{+0.07}_{-0.06}$ & $5.5^{+0.2}_{-0.2}$ & $2.75^{+0.06}_{-0.07}$ & $0.01^{+0.04}_{-0.01}$ & $0.01^{+0.06}_{-0.01}$\\
B275     & $0.20^{+0.05}_{-0.04}$ & $7.2^{+0.5}_{-0.5}$ & $3.23^{+0.13}_{-0.13}$ & $0.34^{+0.12}_{-0.14}$ & $1.00^{+0.45}_{-0.43}$\\
B243     & $0.98^{+0.17}_{-0.21}$ & $4.2^{+0.4}_{-0.2}$ & $2.35^{+0.11}_{-0.06}$ & $0.24^{+0.16}_{-0.08}$ & $0.47^{+0.34}_{-0.18}$\\
B268     & $0.74^{+0.13}_{-0.12}$ & $4.5^{+0.3}_{-0.2}$ & $2.47^{+0.08}_{-0.07}$ & $0.32^{+0.21}_{-0.06}$ & $0.74^{+0.52}_{-0.18}$\\
B86      & $2.24^{+0.05}_{-0.04}$ & $3.1^{+0.0}_{-0.0}$ & $2.00^{+0.01}_{-0.00}$ & $0.15^{+0.02}_{-0.05}$ & $0.21^{+0.04}_{-0.08}$\\
\hline
\end{tabular}
\tablefoot{
For the stars consistent with being on the main sequence, only a lower limit on the age is given and $\varv_{\rm ZAMS}\sin i / \varv_{\rm crit} = \varv\sin i / \varv_{\rm crit}$. The critical rotation velocity, $\varv_{\rm crit}$, is based on the evolutionary mass of the star.}
\end{table*}

\subsection{Surface rotation} \label{p3:sec:rotation}

We did not include a macroturbulent velocity profile in our analysis. As a result, the rotational velocities we find are likely upper limits.
The young age of M17 allows us to determine the spin properties of its (massive) stars close to or at the ZAMS.
Table\,\ref{p3:tab:extra_parameters} lists the present-day projected surface rotation velocity in terms of the critical rotation velocity, $\varv_{\rm crit}$. The latter is determined based on their current radius and mass and is defined as
\begin{equation}
    \varv_{\rm crit} = \sqrt{\frac{G M}{R}}.
\end{equation}
Many of our target stars have yet to reach the ZAMS. To calculate the projected rotation rate at the ZAMS, we contracted the stars from their current radii to their ZAMS radii in accordance with their corresponding MIST evolutionary tracks. 
We assumed that the radial density structure of the stars is represented by the solution of the Lane-Emden equation with a polytropic index $n = 1.5$. Furthermore, we assumed solid-body rotation and angular momentum conservation. The resulting ZAMS radius and projected surface rotation velocity are also listed in Table\,\ref{p3:tab:extra_parameters}. The values for the stars that have already arrived on the main sequence remain unchanged.

\begin{figure}
    \centering
    \includegraphics[width=\columnwidth]{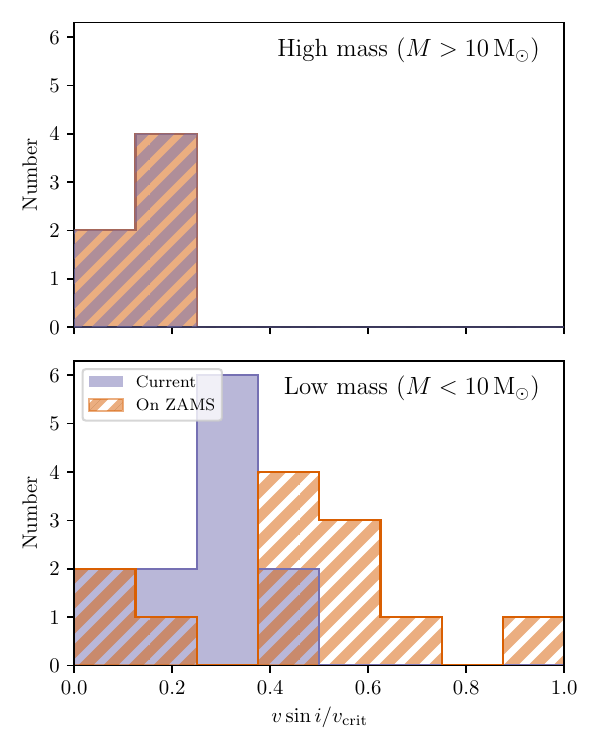}
    \caption{Distribution of the projected equatorial rotational velocities divided by the critical rotation. 
    \emph{Top panel:} Higher-mass stars that have reached the ZAMS. \emph{Bottom panel}: Lower-mass stars that have not yet reached the ZAMS. The value of $\varv \sin i / \varv_{\rm crit}$ is shown for both the current rotation rate and the expected rotation rate when they reach the main sequence. }
    \label{p3:fig:vcrit}
\end{figure}

Figure\,\ref{p3:fig:vcrit} shows the distribution of projected rotation velocity in terms of the critical rotation velocity for both current radii (in orange) and expected ZAMS radii (in purple). For the current radii, we used the measured values as listed in \cref{p3:tab:extra_parameters}, which agree closely with their corresponding evolutionary radii. The sample is split into two parts. In the top panel, the six stars that are more massive than $10\,M_{\odot}$ are shown. All of these have arrived on the ZAMS, with only B181 expected to do so within as little as several tens of thousands of years. However, as the confidence interval of the position of B181 overlaps the ZAMS, we considered it as a main-sequence star. In the bottom panel, the 12 stars that are less massive than $10\,M_{\odot}$ are presented. They have not yet reached the main sequence, and further spin-up because of Kelvin-Helmholtz contraction is relevant for them. The $\varv_{\rm ZAMS}$ distributions of both sets are strikingly different. The massive stars have a velocity distribution up to $\sim 0.3\,\varv_{\rm crit}$ and the intermediate-mass star distribution has velocities up to $\sim 0.6\,\varv_{\rm crit}$. Similar differences exist between the spin distributions of well-established main-sequence O-type \citep{2013A&A...560A..29R,2022A&A...665A.150H} and B-type \citep{2013A&A...550A.109D,2010ApJ...722..605H} stars. This may suggest differences in the mechanism through which these two groups of stars gain angular momentum during their formation. Although the sample is small, the fairly low spin rates of the O-type stars at birth is consistent with the hypothesis that the high-velocity tail of spin rates reported in large studies of presumed-to-be-single main-sequence O-star populations \citep{2013A&A...560A..29R} is actually due to binary interaction, which results in spun-up systems for which the mass donor cannot be detected \citep{2013ApJ...764..166D,2015A&A...580A..92R}.

In the intermediate-mass sample, the sources with a gaseous disk (B243, B268, and B275) in particular will spin up considerably. 
B275 stands out in that it shows a projected equatorial rotation velocity equal to the critical velocity after contraction to the ZAMS. This is possible, but rare, because the star would have to be seen exactly edge on for it not to exceed the critical spin. However, B275 has a gaseous circumstellar disk and is still relatively far from the ZAMS. 
Therefore, we should be cautious to assume that the stellar angular momentum will remain conserved, as interaction with its disk may still alter the stellar angular momentum. 
These same caveats also apply to B243 and B268, which are also relatively distant from the ZAMS and have hot circumstellar gaseous disks. 

\subsection{Comparison with \RT} \label{p3:sec:RT_comparison}

\RT\ performed a similar study of the young stellar content of M17, including a {\sc Fastwind/GA} fit for nine stars. Our sample of 18 stars includes all of these nine objects. A point to note is that we adopted the {\em Gaia} distance of $1675^{+19}_{-18}$\,pc from \citet{2024A&A...681A..21S}, while \RT\ used $1980^{+140}_{-120}$\,pc from \citet{2011ApJ...733...25X}. By itself, this decreases the luminosity of the sources by about 30\%.

Although the spectral analysis takes the same approach, important differences exist. First, the quality of our spectra is superior as we typically stacked three spectra of the quality used by \RT. This allowed us, second, to include abundances of CNO and Si, which are also diagnostics of photospheric properties. Third, RT17 used a bolometric correction on the extinction-corrected absolute $V$ magnitude, whereas we inferred the radius of the star from anchoring our synthetic SED with the absolute $J$ magnitude (given in \cref{p3:tab:extinction_results}).

We limit the comparison to pointing out systematic trends and do not detail the differences for each of the sources separately. For all stars, our $T_{\rm eff}$ values are consistent within the uncertainties with those from \RT. Our luminosities are a factor of two lower on average, with the smallest change being $\sim 10$\,percent and the largest a factor of about four. This results in lower evolutionary masses. Only for B215 do we find a higher luminosity, mainly because of the significantly higher extinction ($A_V = 9.14$ versus $A_V = 7.6$). The extinction to this source is highly uncertain due to limited photometry, however (see Section\,\ref{p3:sec:extinction_discussion} and Appendix\,\ref{p3:sec:app_extinction}). This is further enhanced by the higher temperature found in our analysis (29\,000\,K in our study versus 23\,500\,K by \RT). The improved data quality has been particularly beneficial for the determination of the surface gravity, resulting in considerably smaller uncertainties. This leads to an improved correspondence between the spectroscopic mass, that is, the mass resulting from the gravity and stellar radius, and the evolutionary mass, that is, from a comparison with the MIST evolutionary tracks. A comparison is presented in Appendix\,\ref{p3:app:mass_discrepancy}.

We find projected equatorial rotation velocities that are systematically higher by about 30\,km\,s$^{-1}$. The higher rotation rates could be the result of stacking spectra for which the radial velocity has not been properly corrected. However, the typical radial velocity shifts are up to 20\,km\,s$^{-1}$, which is smaller than the systematic difference we find.

The difference in $\varv \sin i$ is significantly larger than 30\,km\,s$^{-1}$ for B268 and B275. This is likely due to circumstellar and interstellar contamination. The contamination was clipped from the data, resulting in a gap in the center of many line profiles that hampers the determination of the rotation. The impact of this clipping is stronger for the data that were available to \RT. 
The visual extinction $A_{V}$ agrees well for all sources except for B215, as mentioned above. The $R_V$ values are also consistent between this work and \RT, but the uncertainties are relatively large and the scatter between the two works is large as well.

\subsection{Abundances} \label{p3:sec:abundance_discussion}

This study has not been optimized for determining surface abundances. However, we treated the abundances as free parameters to aid the fitting process and constrain them where possible. As pointed out in \cref{p3:sec:results_abundance}, for most stars, the abundance constraints are poor, with the exception of B311. For this O8.5 Vz star, a high signal-to-noise spectrum is combined with a slow projected surface rotation of $\varv \sin i = 30^{+15}_{-25}$\,km\,s$^{-1}$, which reveals many distinct narrow lines. This is beneficial for abundance determinations (see Fig.~\ref{p3:fig:fit_summary_B311}).

The resulting abundances for this star as listed in Table\,\ref{p3:tab:fit_results} all agree within 0.1 dex with the present-day solar photosphere abundances \citep{2009ARA&A..47..481A}. 
This supports the hypothesis that the star and its siblings are young objects. We remark, however, that the present-day composition of the M17 star-forming gas might be expected to be somewhat more metal rich relative to solar as a result of cosmic enrichment over the past 4.6\,Gyr (see, e.g., \citet{2020ApJ...900..179K}).

\subsection{Extinction} \label{p3:sec:extinction_discussion}

\begin{figure}
    \centering
    \includegraphics[width=\columnwidth]{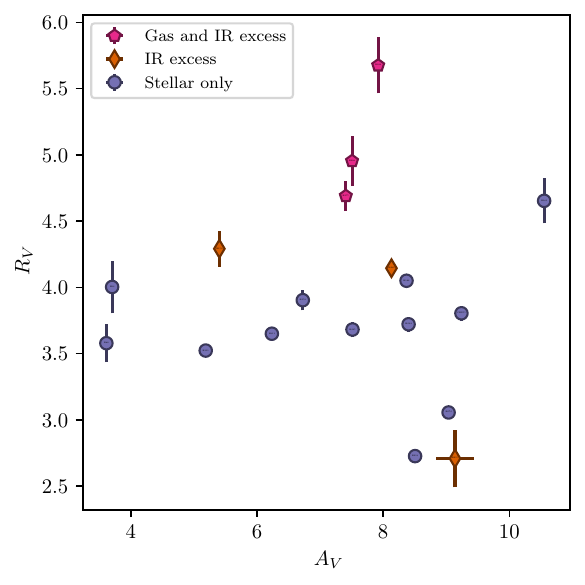}
    \caption{Extinction parameters $A_V$ and $R_V$ of the sources in M17. The markers are identical to those in Figure\,\ref{p3:fig:HRD}. The $R_V$ values tend to be higher for objects with IR excess (with the exception of B215, which has poorly constrained extinction properties). }
    \label{p3:fig:extinction}
\end{figure}

Figure\,\ref{p3:fig:extinction} shows the total visual extinction $A_V$ plotted against total-to-selective extinction $R_V$. The visual extinction toward the sources in M17 varies strongly, with $A_V$ values ranging from 3.6 to 10.6 magnitudes. The range in $R_V$ varies from 2.7 to as high as 5.7. No correlation with the spatial location in M17 of either parameter was found, suggesting that the bulk of the extinction is local to the sources. Moreover, no clear trend of $A_V$ with mass or evolutionary stage was identified. This leaves stochastic line-of-sight differences, that is, some sources lie farther back and/or are more deeply embedded in the natal cloud, to explain the scatter in visual extinction.

The total-to-selective extinction is thought to probe the size distribution and composition of the solid-state material \citep[e.g.,][]{2003ARA&A..41..241D,2014MNRAS.437.1636H,2017ApJ...835..107X}. There is a trend for $R_{V}$ values to be highest for objects with IR excess. 
This suggests that the material causing the extinction in these sources is nearby, for instance, in the circumstellar disk or in the direct ambient medium from which the disks have been accreting, because the processing of the grain material that causes the higher $R_V$ is otherwise not expected to be markedly different from that of the other sources.
An exception to the trend is B215, which has relatively poorly constrained extinction properties. 

\subsection{Mass-loss rates} \label{p3:sec:mass_loss_rates}

\begin{figure}
    \centering
    \includegraphics[width=\columnwidth]{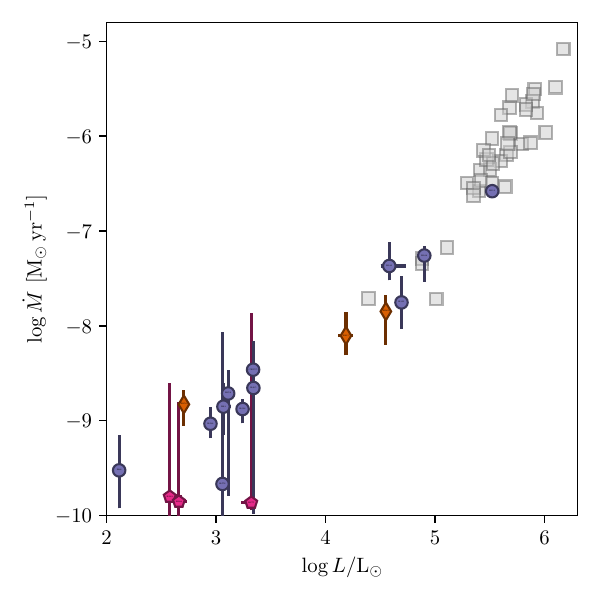}
    \caption{Mass-loss rate as a function of luminosity for the M17 sample. The markers are identical to those in Fig.~\ref{p3:fig:HRD}. The additional sources shown as gray  squares are the Galactic OB stars from \citet{2004A&A...415..349R}, \citet{2005A&A...441..711M}, and \citet{2006A&A...446..279C}. }
    \label{p3:fig:mass_loss}
\end{figure}

Mass-loss sensitive lines available in our optical spectra are H$\alpha$, to some extent H$\beta$, and for O-type stars, \ion{He}{ii}\,4686. 
The determination of the mass-loss rates from H$\alpha$ and H$\beta$ is hindered by contamination by interstellar and, if present, circumstellar emission. Because of interstellar emission, the line centers, which are most sensitive to mass loss, had to be clipped. An additional complication is the potential contamination of the broad wings of the lines from circumstellar material that has not been clipped. This broad nonstellar emission could originate from a disk or disk wind. If this wind emission is present, the mass-loss rate may be overestimated because the circumstellar emission is fit as if it originated in the stellar winds.

For the hottest and brighter stars ($\log L / L_\odot > 4$), we can place tight constraints on $\dot{M}$ based on our assumption on $\beta$. Reasonable variations of $\beta$ could change the determined mass-loss rate by up to a factor 2 \citep{2004A&A...413..693M}. For some of the fainter stars without disk signatures, we also constrained the mass-loss rates, to the best of our knowledge, for the first time in this part of parameter space, but as the wind signatures are extremely weak, we remain cautious with respect to the obtained results. The mass-loss determinations for these $11\,000 - 18\,000$\,K and $\sim 10^{2} - 2 \times 10^{3}\,L_{\odot}$ stars should be treated with care. 

Figure\,\ref{p3:fig:mass_loss} shows the mass-loss rate as a function of luminosity, along with $\dot{M}$ determined by \citet{2004A&A...415..349R}, \citet{2005A&A...441..711M}, and \citet{2006A&A...446..279C} for other Galactic data sets. The values derived by these authors assume smooth winds; we corrected their rates adopting the clumping factor ($f_{\rm cl} = 10$) used in our analysis. We find the mass-loss rates of the brightest stars in our sample are consistent with these previous studies of Galactic O- and early B-type stars.

For stars without disk signatures that are dimmer than $2 \times 10^{3}\,L_{\odot}$, we find mass-loss rates (in M$_{\odot}$\,yr$^{-1}$) up to $\log\dot{M}\sim-8.5$. Although these are extremely low rates, they show that the initiation of the stellar outflow already occurs during the PMS phase. Assuming the wind-driving mechanism is radiation pressure on spectra lines, \citet{2014A&A...564A..70K} predicted mass-loss rates for main-sequence B-type stars in this part of parameter space (their models T14, T16, T18, and T20). Typical $\dot{M}$ values are $\sim\,10^{-11}\,M_{\odot}$\,yr$^{-1}$, that is, up to about two orders of magnitude lower than we find. Taking our results at face value, this either indicates that the theoretical predictions are significantly too low, or that PMS B stars have considerably stronger mass loss than main-sequence B stars in the same part of the ($T_{\rm eff},L$) space. As pointed out above, for the PMS stars with disks, we consider the mass-loss rates as tentative upper limits.

\section{Conclusions} \label{p3:sec:conclusion}

We have analyzed optical spectra of 18 stars in the young star-forming region M17 with spectral types ranging from O4.5 to B9, corresponding to temperatures from 46\,000 to 11\,000\,K, using the stellar atmosphere model {\sc Fastwind} and fitting algorithm {\sc Kiwi-GA}. The sources are still highly reddened, with visual extinctions between $A_V=3.6$ and 10.6 mag. Three of the stars show gaseous and dusty circumstellar material that is likely to be remnant features of their formation. Three other stars display hot dusty circumstellar material in their SEDs. 
The sample is unique in that it constitutes a population of massive and intermediate-mass stars close to or on the zero-age main sequence, allowing us to study the actual outcome of star formation in this mass range. Our main conclusions are listed below.

\begin{itemize}

    \item The HRD positions of the six most massive stars (${13 \lesssim M \lesssim 47}$\,M$_\odot$) are consistent with the theoretical location of the zero-age main sequence as given by the MIST evolutionary tracks; the 12 lower-mass stars (${3 \lesssim M \lesssim  7\,}$M$_{\odot}$) are PMS sources. The cluster age is constrained to $0.4_{-0.2}^{+0.6}$\,Myr based on the PMS population. We conclude that high-mass ZAMS stars can be found by searching for highly reddened sources in star-forming regions.

    \item The three sample stars with gaseous disks (B275, B243, and B286) are located farther from the main sequence than the stars without a disk, which is in line with the hypothesis that these are the least evolved objects in the sample and have not yet lost their accretion disk.
    \item Within the short time span of the formation process of these stars, we find a strong correlation between age and stellar mass, with lower-mass stars being older. This is inconsistent with these stars crossing the birthline simultaneously during formation. 

    \item We identify a dichotomy between the projected equatorial rotational velocity for stars $M > 13\,M_{\odot}$, which are O-type stars on the ZAMS, and stars $3 \lesssim M \lesssim 7\,M_{\odot}$, which will become B-type stars on the ZAMS. When we extrapolate the projected spin velocities to the ZAMS for all stars, the high-mass stars have $\varv \sin i \lesssim 0.3\,\varv_{\rm crit}$ and the intermediate-mass stars have $\varv \sin i \lesssim 0.6\,\varv_{\rm crit}$. This is in line with previous results for well-established main-sequence populations and may suggest different mechanisms for angular momentum accretion between the two groups.
    
    \item Surface abundances of C, N, O, and Si are most accurate for the slowly spinning and bright source B311. They match those of the current solar abundances, which supports the young age of M17 and suggests that the gas from which the cluster formed has not been significantly chemically enriched relative to the solar nebula gas.

    \item No strong spatial pattern of extinction properties can be identified. We do find a correlation between $R_V$ and the presence or absence of circumstellar matter. The stars showing signs of circumstellar disks have higher $R_V$ values, suggesting that the different size and composition properties of the solid-state material in the line of sight is due to the local conditions close to the star.

    \item{For the stars more massive than 10\,M$_{\odot}$ we find mass-loss rates that are in line with the observed rates of Galactic main-sequence O\,V-III stars. For the first time, we derive mass-loss properties for PMS B\,V-III stars and find rates of up to $10^{-8.5}\,$M$_{\odot}$\,yr$^{-1}$. These should be considered as tentative, but establish at face value that stellar winds are already initiated in the PMS phase. They are higher by two orders of magnitude than the predictions for main-sequence B stars by \citet{2014A&A...564A..70K}. If these theoretical rates are correct, PMS-star winds might be considerably stronger than the MS-star winds of stars with similar stellar properties.}

\end{itemize}

This unique sample of stars gives insight into the outcome of massive star formation, providing both constraints for star formation theory and initial conditions for stellar evolution. The mass-loss properties of the PMS population remain uncertain, but could be improved with more focused studies. More data on the wind diagnostics and the contaminants could benefit studies like this. UV detections of the outflows would be very insightful, but are extremely difficult to obtain because the extinction is strong. Alternatively, James Webb Space Telescope observations of the more sensitive Br$\alpha$ might be able to provide improved mass-loss estimates \citep{2008A&ARv..16..209P,2011A&A...535A..32N}.

\begin{acknowledgements}
We thank the referee for insightful comments that helped to improve this manuscript. This publication is part of the project ‘Massive stars in low-metallicity environments: the progenitors of massive black holes’ with project number OND1362707 of the research TOP-programme, which is (partly) financed by the Dutch Research Council (NWO). FB acknowledges the support of the European Research Council (ERC) Horizon Europe under grant agreement number 101044048. MCRT acknowledges support by the German Aerospace Center (DLR) and the Federal Ministry for Economic Affairs and Energy (BMWi) through program 50OR2314 ‘Physics and Chemistry of Planet-forming disks in extreme environments’. This work is based on observations collected at the European Organization for Astronomical Research in the Southern Hemisphere under ESO programs 60.A-9404(A), 085.D-0741, 089.C-0874(A), 091.C-0934(B) and 103.D-0099. We thank SURF (www.surf.nl) for the support in using the National Supercomputer Snellius. This research has made use of NASA’s Astrophysics Data System.  This work makes use of the Python programming language\footnote{Python Software Foundation; \url{https://www.python.org/}}, in particular packages including NumPy \citep{harris2020array}, SciPy \citep{2020SciPy-NMeth}, and Matplotlib \citep{Hunter:2007}.

\end{acknowledgements}

\bibliographystyle{aa} 
\bibliography{aanda}

\begin{appendix}

\begin{landscape}

\section{Fit results} \label{p3:app:results}

\begin{table}[h!]
\caption{Best fit stellar parameters from {\sc Fastwind}/GA fitting.}
\label{p3:tab:fit_results}
\def\arraystretch{1.5}
\tiny
\begin{tabular}{lccccccccccccccc} \hline \hline
Source & $T_{\rm eff}$ & $\log g$ & $\log \dot{M}$ & $\varv\sin i$ & $y_{\rm He}$ & $\epsilon_{\rm C}$ & $\epsilon_{\rm N}$ & $\epsilon_{\rm O}$ & $\epsilon_{\rm Si}$ & $\log L$ & $R$ & $M_{\rm spec}$ & $\log Q_0$ \\
& [K] & [cm\,s$^{-2}$] & [M$_\odot$\,yr$^{-1}$] & [km\,s$^{-1}$] & [$n_{\rm He} / n_{\rm H}$] & [$\log n_{\rm C} / n_{\rm H} + 12$] & [$\log n_{\rm N} / n_{\rm H} + 12$] & [$\log n_{\rm x} / n_{\rm O} + 12$] & [$\log n_{\rm Si} / n_{\rm H} + 12$] & [L$_\odot$] & [R$_\odot$] & [M$_\odot$] & [s$^{-1}$] \\ \hline 
B111 & $45750^{+1250}_{-2250}$ & $3.92^{+0.08}_{-0.12}$ & $-6.58^{+0.08}_{-0.05}$ & $205^{+30}_{-20}$ & $0.09^{+0.03}_{-0.01}$ & $8.7^{+0.3}_{-0.6}$ & $8.2^{+0.2}_{-0.6}$ & $8.8^{+0.2}_{-2.8}$ & $6.2^{+2.1}_{-0.2}$ & $5.52^{+0.04}_{-0.06}$ & $9.25^{+0.37}_{-0.30}$ & $26.0^{+3.7}_{-5.2}$ & $49.32^{+0.05}_{-0.08}$ \\ 
B164 & $39000^{+1250}_{-500}$ & $4.02^{+0.18}_{-0.12}$ & $-7.26^{+0.10}_{-0.28}$ & $120^{+15}_{-15}$ & $0.09^{+0.03}_{-0.01}$ & $8.1^{+0.6}_{-0.3}$ & $7.8^{+0.8}_{-1.1}$ & $8.8^{+0.2}_{-1.0}$ & $7.7^{+0.5}_{-0.8}$ & $4.90^{+0.04}_{-0.03}$ & $6.24^{+0.19}_{-0.21}$ & $14.9^{+6.1}_{-3.2}$ & $48.50^{+0.08}_{-0.04}$ \\ 
B311 & $36250^{+500}_{-1250}$ & $4.24^{+0.16}_{-0.22}$ & $-7.75^{+0.28}_{-0.28}$ & $30^{+15}_{-25}$ & $0.08^{+0.02}_{-0.01}$ & $8.4^{+0.3}_{-0.5}$ & $7.8^{+0.5}_{-0.5}$ & $8.7^{+0.3}_{-0.5}$ & $7.5^{+0.3}_{-0.6}$ & $4.69^{+0.03}_{-0.04}$ & $5.68^{+0.21}_{-0.18}$ & $20.4^{+7.9}_{-7.3}$ & $48.09^{+0.06}_{-0.12}$ \\ 
B181 & $29500^{+4250}_{-2000}$ & $3.60^{+0.28}_{-0.26}$ & $-7.37^{+0.25}_{-0.15}$ & $180^{+30}_{-25}$ & $0.09^{+0.06}_{-0.01}$ & $7.8^{+0.6}_{-1.3}$ & $6.9^{+2.2}_{-0.9}$ & $7.7^{+0.8}_{-1.6}$ & $7.6^{+0.5}_{-1.6}$ & $4.58^{+0.15}_{-0.08}$ & $7.55^{+0.39}_{-0.62}$ & $8.3^{+6.5}_{-3.6}$ & $47.33^{+0.69}_{-0.45}$ \\ 
B289 & $34250^{+1500}_{-750}$ & $3.92^{+0.20}_{-0.16}$ & $-7.85^{+0.18}_{-0.35}$ & $180^{+20}_{-20}$ & $0.09^{+0.03}_{-0.01}$ & $8.3^{+0.3}_{-0.6}$ & $7.6^{+0.8}_{-1.6}$ & $8.3^{+0.7}_{-2.4}$ & $7.7^{+0.3}_{-0.8}$ & $4.55^{+0.05}_{-0.03}$ & $5.40^{+0.18}_{-0.21}$ & $8.8^{+4.4}_{-2.5}$ & $47.86^{+0.14}_{-0.07}$ \\ 
B215 & $29000^{+1600}_{-1400}$ & $4.18^{+0.16}_{-0.18}$ & $-8.10^{+0.25}_{-0.20}$ & $245^{+25}_{-30}$ & $0.09^{+0.03}_{-0.01}$ & $8.0^{+0.4}_{-0.5}$ & $8.6^{+0.6}_{-2.4}$ & $8.1^{+0.6}_{-0.8}$ & $6.9^{+0.7}_{-0.9}$ & $4.19^{+0.07}_{-0.07}$ & $4.95^{+0.27}_{-0.28}$ & $13.5^{+5.2}_{-4.4}$ & $46.58^{+0.30}_{-0.26}$ \\ 
B93 & $17800^{+1200}_{-200}$ & $3.92^{+0.18}_{-0.06}$ & $-8.85^{+0.25}_{-0.30}$ & $260^{+20}_{-25}$ & $0.11^{+0.02}_{-0.03}$ & $7.7^{+0.5}_{-0.6}$ & $7.0^{+2.5}_{-0.9}$ & $9.1^{+0.3}_{-2.9}$ & $6.9^{+0.2}_{-0.6}$ & $3.07^{+0.07}_{-0.03}$ & $3.62^{+0.12}_{-0.17}$ & $4.0^{+1.4}_{-0.4}$ & $43.36^{+0.21}_{-0.04}$ \\ 
B205 & $17200^{+400}_{-1000}$ & $3.82^{+0.12}_{-0.20}$ & $-8.66^{+0.30}_{-0.25}$ & $200^{+30}_{-30}$ & $0.09^{+0.02}_{-0.01}$ & $7.4^{+0.5}_{-1.4}$ & $7.4^{+1.4}_{-1.4}$ & $9.3^{+0.2}_{-1.1}$ & $7.1^{+0.4}_{-0.6}$ & $3.34^{+0.03}_{-0.06}$ & $5.32^{+0.26}_{-0.18}$ & $6.8^{+1.8}_{-2.1}$ & $43.50^{+0.05}_{-0.26}$ \\ 
CEN55 & $18000^{+1200}_{-600}$ & $3.74^{+0.06}_{-0.16}$ & $-8.46^{+0.30}_{-1.52}$ & $5^{+20}_{-5}$ & $0.19^{+0.01}_{-0.03}$ & $8.1^{+0.1}_{-0.3}$ & $9.4^{+0.1}_{-3.5}$ & $9.3^{+0.2}_{-0.8}$ & $7.0^{+0.3}_{-0.2}$ & $3.34^{+0.07}_{-0.04}$ & $4.84^{+0.19}_{-0.25}$ & $4.7^{+0.6}_{-1.3}$ & $43.80^{+0.22}_{-0.17}$ \\ 
B234 & $16600^{+800}_{-200}$ & $3.82^{+0.14}_{-0.12}$ & $-9.03^{+0.18}_{-0.15}$ & $350^{+40}_{-25}$ & $0.10^{+0.03}_{-0.02}$ & $7.2^{+0.6}_{-0.2}$ & $7.5^{+1.9}_{-0.9}$ & $8.0^{+2.1}_{-1.0}$ & $7.0^{+0.6}_{-0.1}$ & $2.95^{+0.05}_{-0.03}$ & $3.64^{+0.12}_{-0.15}$ & $3.2^{+0.9}_{-0.7}$ & $42.92^{+0.20}_{-0.03}$ \\ 
B213 & $17250^{+250}_{-1000}$ & $3.68^{+0.10}_{-0.18}$ & $-8.88^{+0.10}_{-0.15}$ & $245^{+20}_{-20}$ & $0.10^{+0.04}_{-0.02}$ & $7.7^{+0.3}_{-0.6}$ & $8.4^{+0.9}_{-2.5}$ & $6.1^{+2.7}_{-0.1}$ & $7.0^{+0.6}_{-0.2}$ & $3.24^{+0.03}_{-0.06}$ & $4.72^{+0.21}_{-0.14}$ & $3.9^{+0.8}_{-1.0}$ & $43.43^{+0.09}_{-0.23}$ \\ 
B253 & $15400^{+600}_{-800}$ & $3.86^{+0.12}_{-0.14}$ & $-8.71^{+0.25}_{-1.08}$ & $170^{+60}_{-40}$ & $0.09^{+0.02}_{-0.01}$ & $7.7^{+0.4}_{-1.8}$ & $7.0^{+2.5}_{-1.1}$ & $6.8^{+2.2}_{-0.8}$ & $6.8^{+0.3}_{-0.5}$ & $3.11^{+0.04}_{-0.05}$ & $5.09^{+0.22}_{-0.19}$ & $6.8^{+1.5}_{-1.5}$ & $42.81^{+0.16}_{-0.21}$ \\ 
B150 & $15600^{+800}_{-800}$ & $3.80^{+0.18}_{-0.14}$ & $-8.83^{+0.15}_{-0.23}$ & $315^{+25}_{-30}$ & $0.14^{+0.04}_{-0.03}$ & $8.1^{+0.3}_{-0.9}$ & $6.8^{+3.3}_{-0.3}$ & $9.3^{+0.6}_{-2.4}$ & $7.0^{+0.2}_{-0.1}$ & $2.71^{+0.05}_{-0.05}$ & $3.11^{+0.14}_{-0.14}$ & $2.2^{+0.8}_{-0.5}$ & $42.55^{+0.13}_{-0.21}$ \\ 
B272 & $16000^{+1200}_{-800}$ & $4.68^{+0.18}_{-0.20}$ & $-9.67^{+1.60}_{-0.35}$ & $5^{+25}_{-5}$ & $0.30^{+0.01}_{-0.09}$ & $8.8^{+0.5}_{-1.6}$ & $8.3^{+1.6}_{-1.9}$ & $9.5^{+0.6}_{-2.5}$ & $8.0^{+0.7}_{-0.7}$ & $3.06^{+0.07}_{-0.05}$ & $4.44^{+0.20}_{-0.25}$ & $34.4^{+14.6}_{-11.9}$ & $42.98^{+0.36}_{-0.18}$ \\ 
B275 & $12750^{+750}_{-1250}$ & $3.44^{+0.12}_{-0.34}$ & $-9.87^{+2.00}_{-0.08}$ & $185^{+65}_{-75}$ & $0.10^{+0.10}_{-0.03}$ & $6.3^{+2.0}_{-0.4}$ & $9.3^{+0.2}_{-3.4}$ & $6.6^{+2.5}_{-0.7}$ & $6.7^{+0.6}_{-0.6}$ & $3.32^{+0.05}_{-0.09}$ & $9.46^{+0.72}_{-0.42}$ & $9.0^{+1.9}_{-4.3}$ & $42.38^{+0.26}_{-3.05}$ \\ 
B243 & $11900^{+1400}_{-300}$ & $3.78^{+0.22}_{-0.14}$ & $-9.80^{+1.20}_{-0.20}$ & $140^{+95}_{-45}$ & $0.09^{+0.06}_{-0.01}$ & $8.0^{+0.8}_{-1.1}$ & $9.8^{+0.2}_{-3.2}$ & $7.3^{+2.8}_{-0.3}$ & $7.0^{+0.3}_{-0.1}$ & $2.58^{+0.11}_{-0.03}$ & $4.61^{+0.15}_{-0.40}$ & $4.7^{+1.9}_{-1.1}$ & $41.00^{+0.88}_{-1.68}$ \\ 
B268 & $11300^{+900}_{-300}$ & $3.78^{+0.22}_{-0.12}$ & $-9.86^{+1.05}_{-0.15}$ & $180^{+115}_{-35}$ & $0.10^{+0.08}_{-0.02}$ & $7.8^{+0.7}_{-0.6}$ & $7.0^{+3.0}_{-0.3}$ & $7.0^{+2.4}_{-0.1}$ & $7.1^{+0.5}_{-0.1}$ & $2.66^{+0.07}_{-0.04}$ & $5.63^{+0.22}_{-0.36}$ & $7.0^{+3.2}_{-1.3}$ & $40.95^{+0.33}_{-2.26}$ \\ 
B86 & $11600^{+400}_{-200}$ & $4.02^{+0.10}_{-0.08}$ & $-9.52^{+0.38}_{-0.40}$ & $95^{+15}_{-35}$ & $0.10^{+0.03}_{-0.03}$ & $7.4^{+0.4}_{-0.5}$ & $6.8^{+1.8}_{-0.2}$ & $7.6^{+1.8}_{-0.2}$ & $7.0^{+0.2}_{-0.1}$ & $2.11^{+0.04}_{-0.03}$ & $2.85^{+0.09}_{-0.10}$ & $3.1^{+0.6}_{-0.4}$ & $40.52^{+0.45}_{-0.41}$ \\ 
\hline
\end{tabular}
\end{table}

\end{landscape}

\begin{landscape}

\begin{figure}
    \includegraphics[width=1.3\textheight]{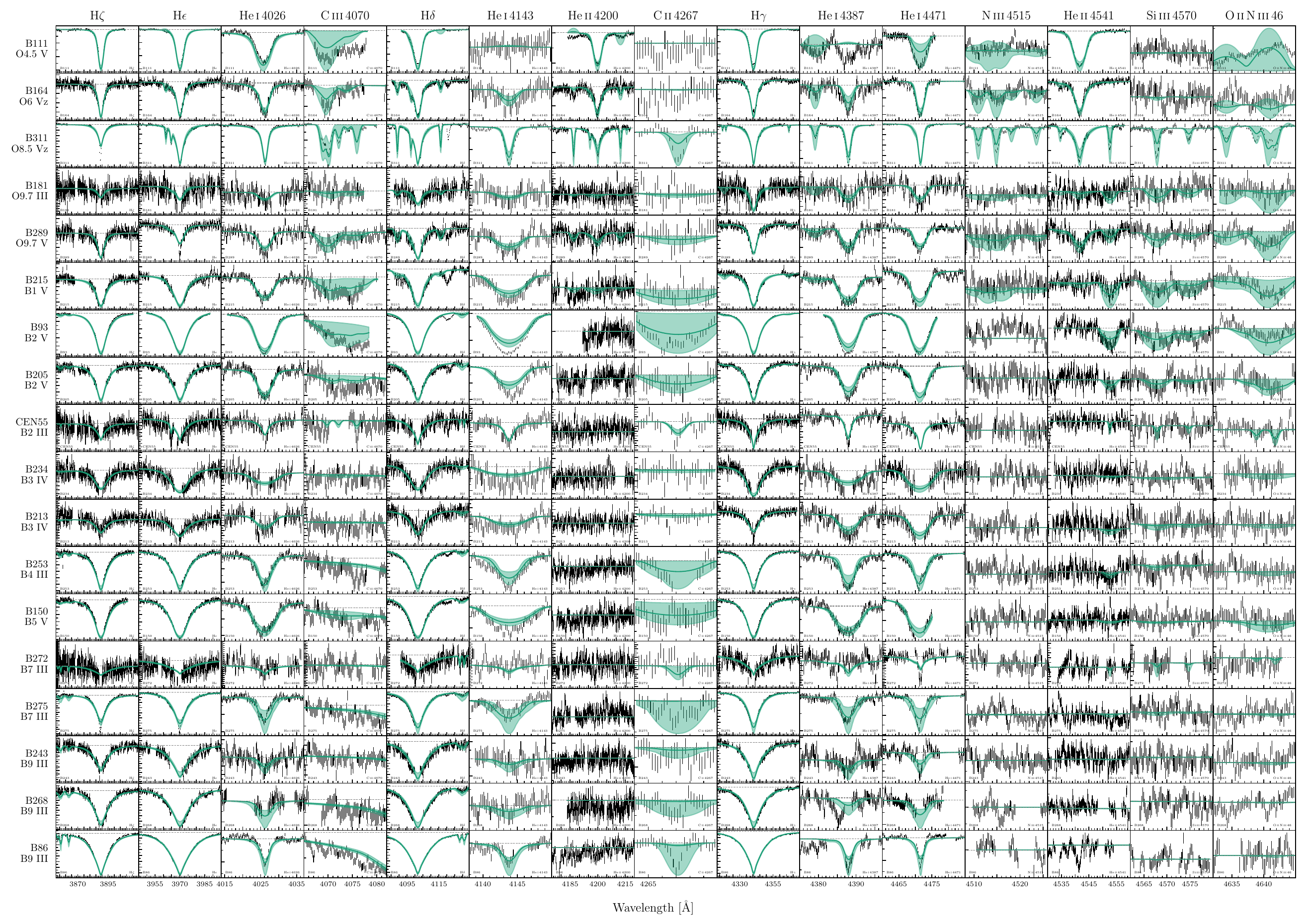}
    \caption{Overview of {\sc Fastwind/Kiwi-GA} fits of all studied lines of each star in the sample. The black vertical bars indicate the observed data, with the length indicating the 1$\sigma$ uncertainty on the data. The green line shows the best fit model and the shaded region the range of models in the 1$\sigma$ confidence interval. A more detailed overview of the fit results is shown in Figure~\ref{p3:fig:fit_summary_B311} for B311 and in \cref{p3:sec:fit_summaries}, but also available online at \url{https://doi.org/10.5281/zenodo.13285505}. }
    \label{p3:fig:fit_overview}
\end{figure}

\end{landscape}

\begin{landscape}
    
\begin{figure}\ContinuedFloat
    \includegraphics[width=1.3\textheight]{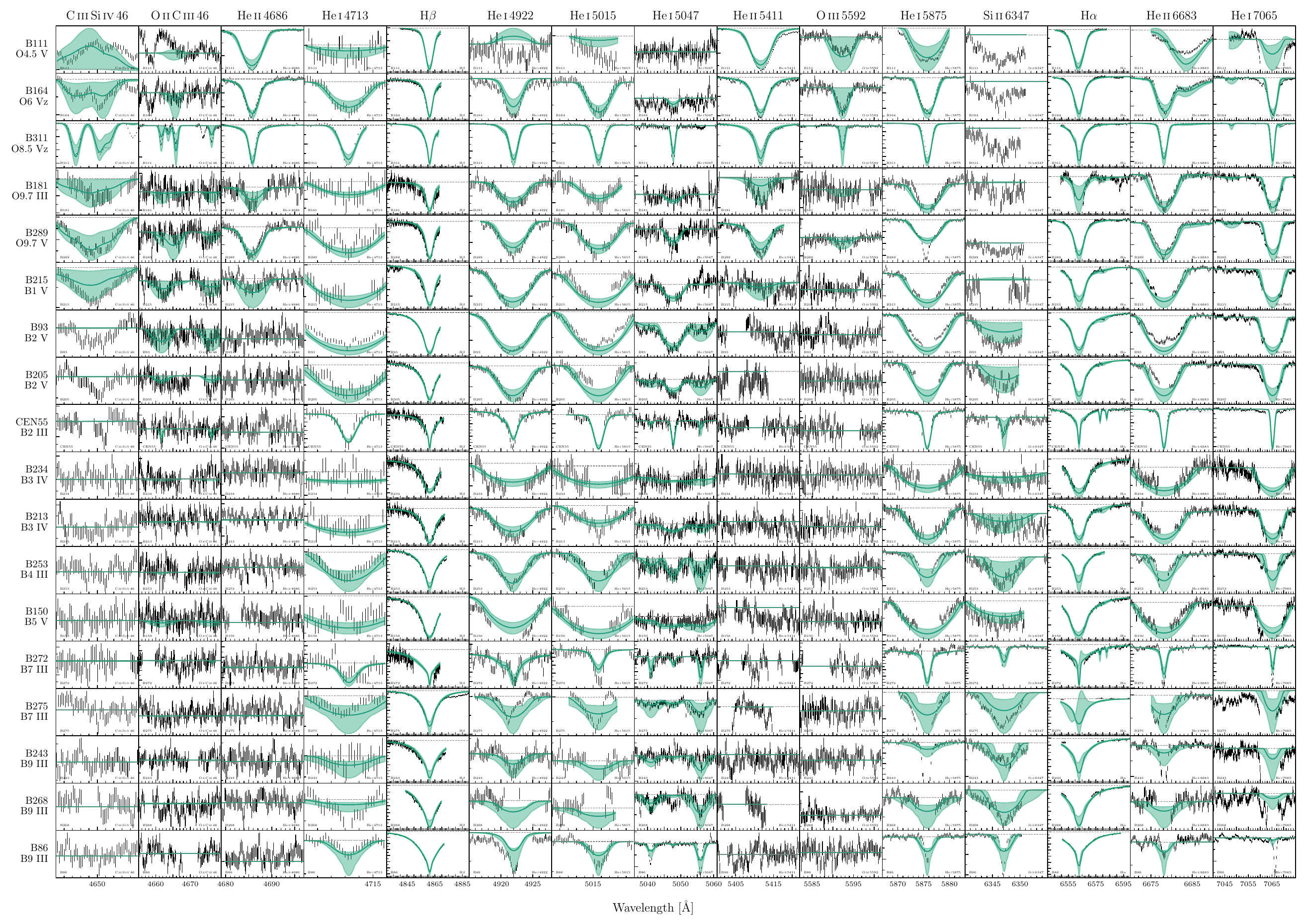}
    \caption{continued. }
\end{figure}

\end{landscape}

\section{Extinction fits} \label{p3:sec:app_extinction}
\begin{figure*}
    \centering
    \includegraphics[width=\textwidth]{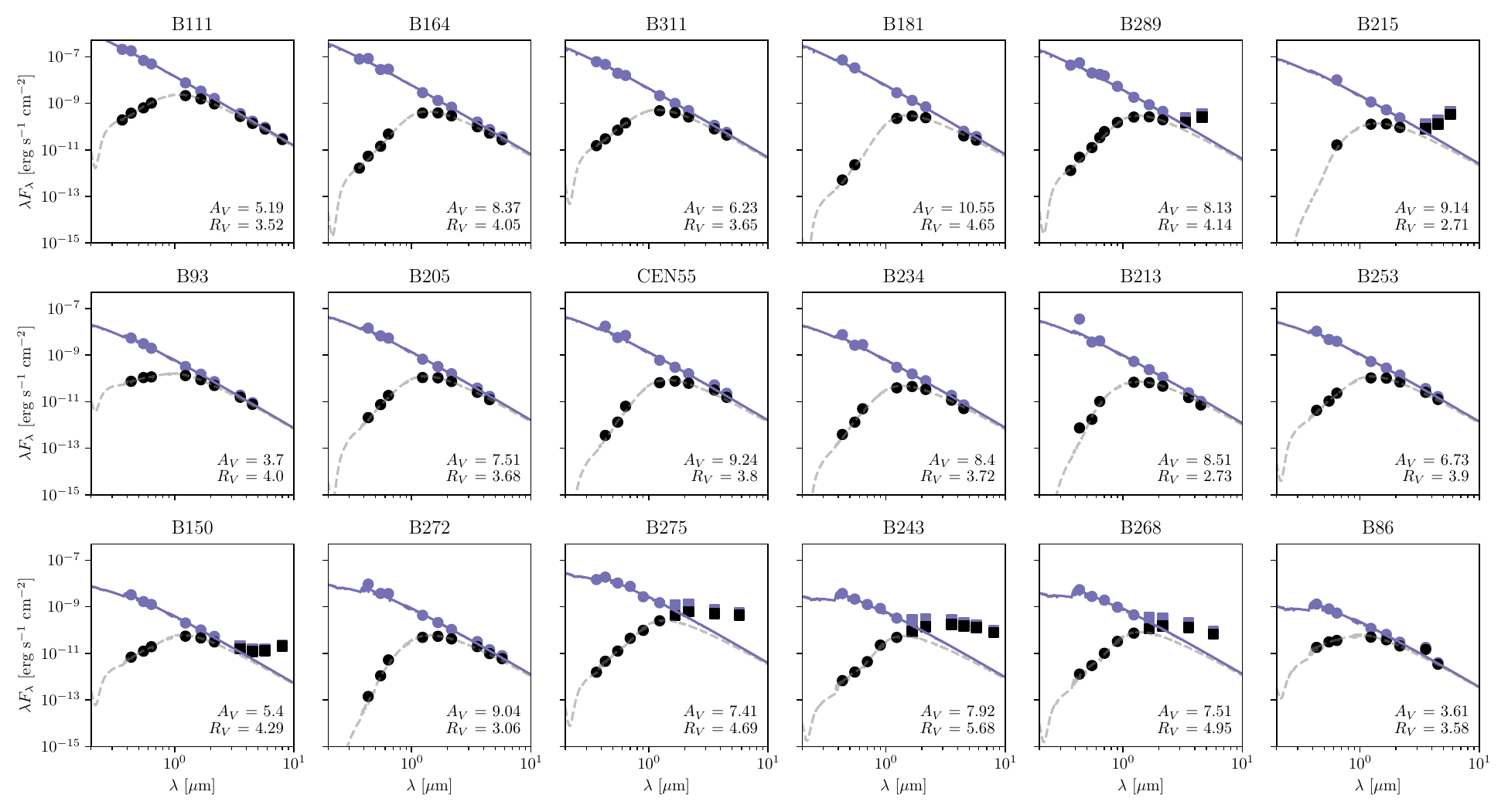}
    \caption{Observed and dereddened SED of each star. Black squares and dots indicate observed photometric data, purple markers indicate dereddened data. The squares indicate data with IR excess emission; these are not used for the determination of the extinction. The dashed gray  line shows the reddened stellar model and the purple solid line the original stellar model.}
    \label{p3:fig:dereddened_SED}
\end{figure*}

\begin{figure*}
    \centering
    \includegraphics[width=\textwidth]{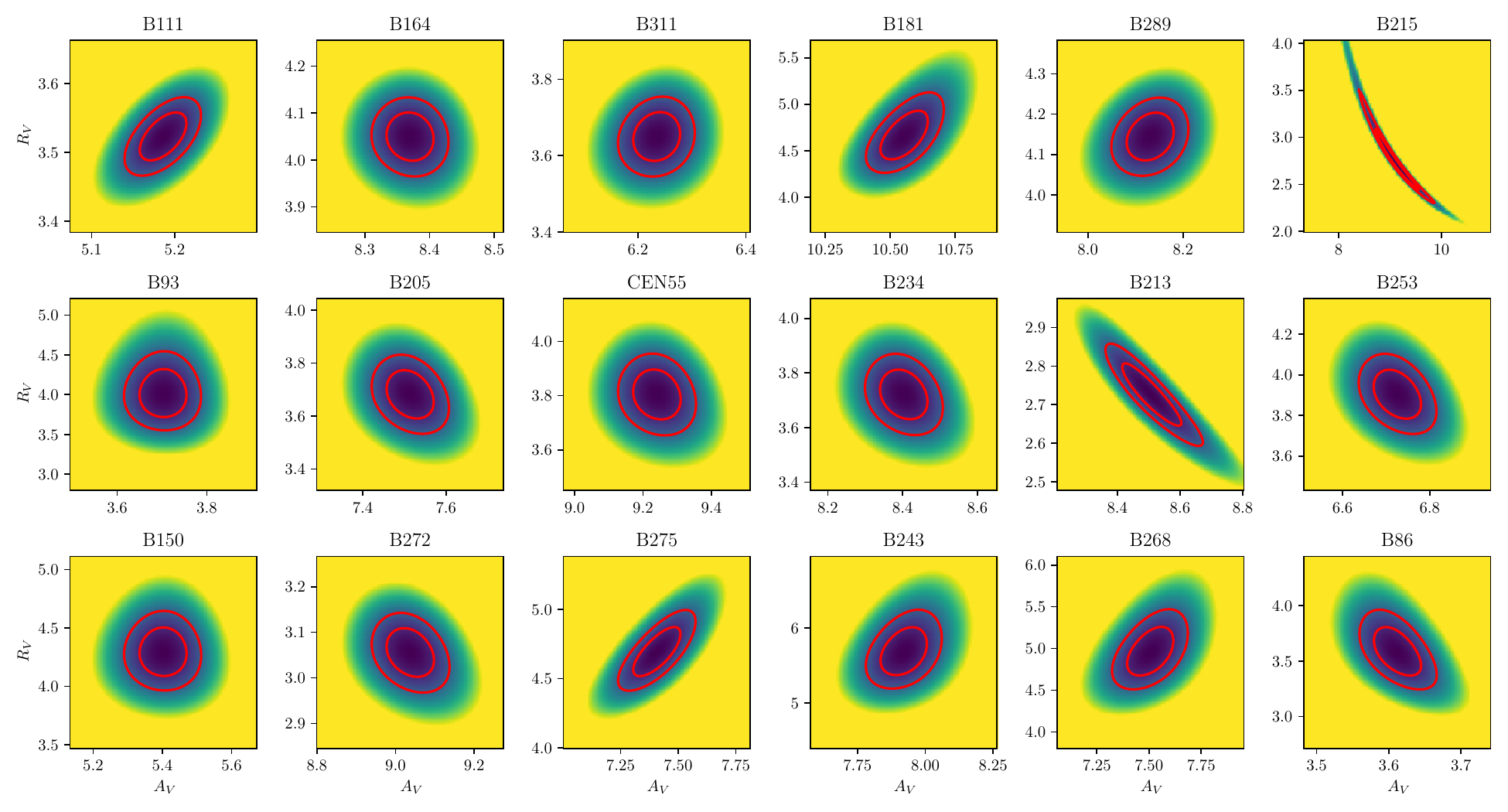}
    \caption{$A_V$ and $R_V$ relations found in the fitting process. The color indicates the $\chi^2$ value of the combination of $A_V$ and $R_V$ and the red lines indicate the 1 $\sigma$ and 2 $\sigma$ confidence contours. }
    \label{p3:fig:AvRv_correlation}
\end{figure*}

The extinction towards each of the stars varies strongly. Figure~\ref{p3:fig:dereddened_SED} shows an overview of the reddened and dereddened SEDs of the stars. The corresponding $A_V$ and $R_V$ values are also indicated. Figure~\ref{p3:fig:AvRv_correlation} shows the relation between $A_V$ and $R_V$ for each star. Some stars show strong relations, this can be due to a small number of photometric data points being available for the fit. B215 has only four data points available for the fit, as the longer wavelength data shows strong IR excess emission. As a result the three free parameters of the fit become quite degenerate and we find a strong relation. However, the resulting uncertainty on $A_V$ and $R_V$, and the absolute magnitude remains limited, see Table~\ref{p3:tab:extinction_results}. The best fit value might be unreliable and subject to significant change if more photometric data are added.  

\FloatBarrier

\section{Mass discrepancy} \label{p3:app:mass_discrepancy}
\begin{figure}
    \centering
    \includegraphics{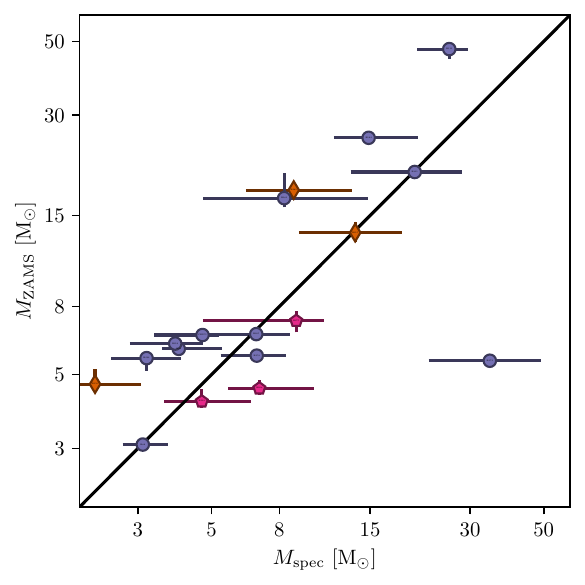}
    \caption{Stellar mass based on MIST evolutionary tracks plotted against mass based on spectroscopy. The correspondence between the two estimates is good and at the low-mass end does not reveal systematic differences. Possibly, the derived evolutionary masses are higher at the high-mass end. Markers are identical to Figure\,\ref{p3:fig:HRD}. }
    \label{p3:fig:mass_discrep}
\end{figure}

\citet{1992A&A...261..209H} first noted the difference between spectroscopic surface gravity of stars and their inferred evolutionary model surface gravity. Since then improved models have decreased the gap, particularly above 15\,M$_\odot$ \citep[e.g][]{2002A&A...396..949H}. However, \citet{2004A&A...415..349R} find a systematic offset for stars above 15\,M$_\odot$. Figure\,\ref{p3:fig:mass_discrep} shows the evolutionary mass at the ZAMS as function of the spectroscopic mass. No clear trend or offset is visible at the low mass end. At higher masses there appears to be a slight trend in which higher evolutionary masses are found. One outlier stands out, which is B272 showing a significantly higher spectroscopic mass than evolutionary mass. This is due to the high surface gravity found. A possible cause of the high value of $\log g$ could be the low signal-to-noise ratio and contamination in the spectrum, resulting in a strange temperature and gravity combination as best fit.  

\FloatBarrier

 \section{Fit summaries}\label{p3:sec:fit_summaries}

\begin{figure*}
     \centering
     \includegraphics[width=0.9\textwidth]{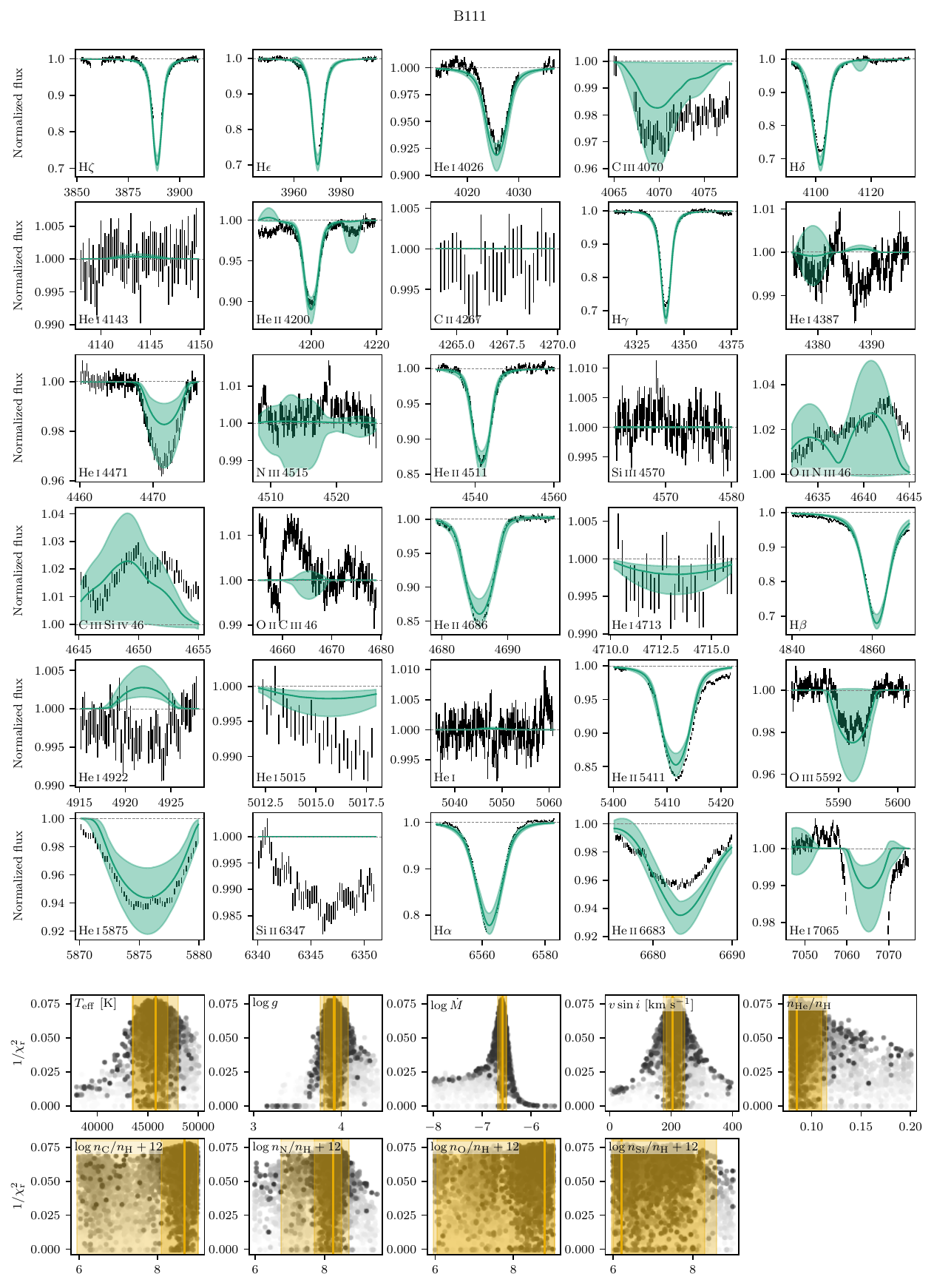}
     \caption{Same as Figure\,\ref{p3:fig:fit_summary_B311}, but for B111.}
 \end{figure*}

 \begin{figure*}
     \centering
     \includegraphics[width=0.9\textwidth]{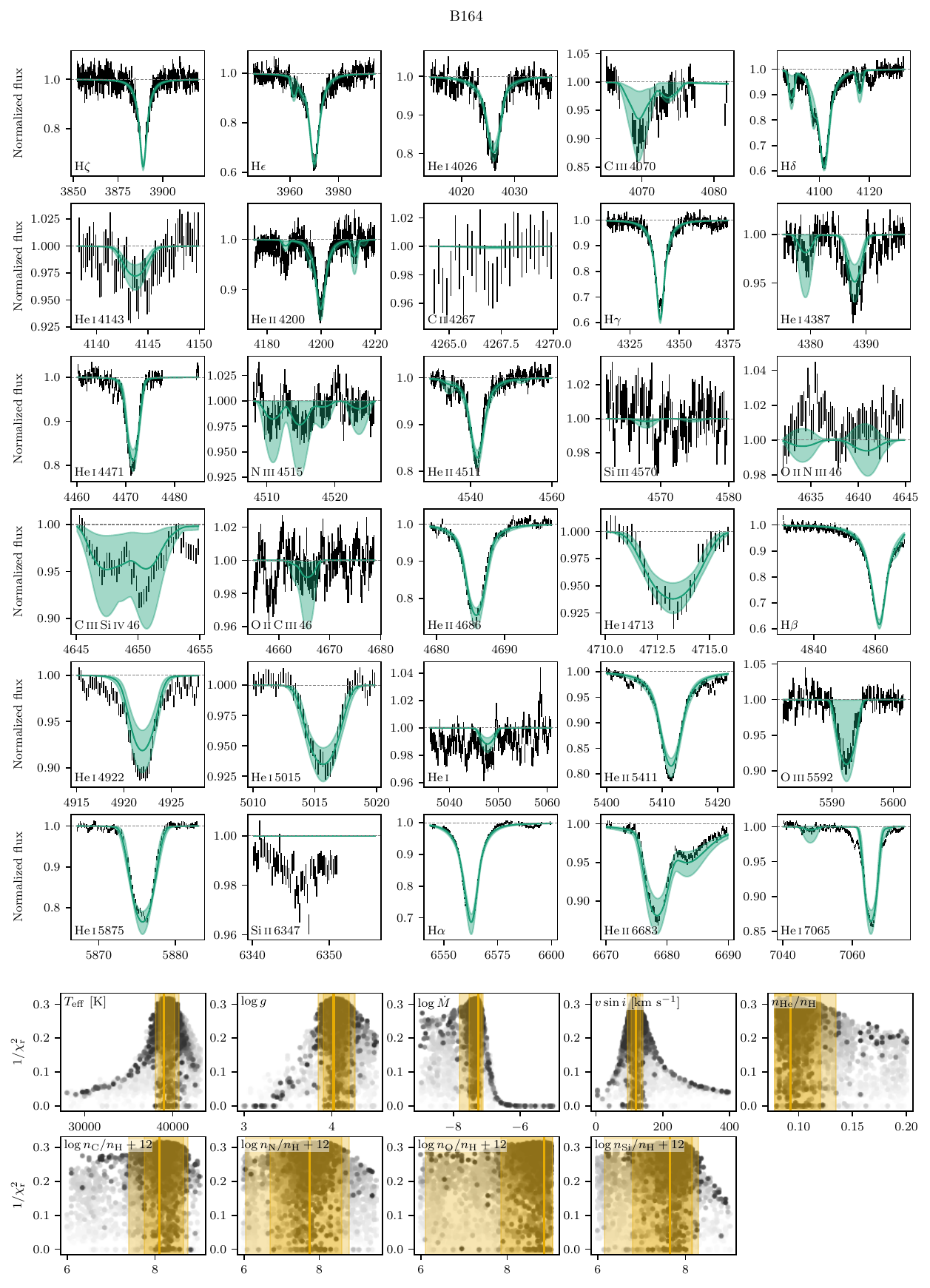}
     \caption{Same as Figure\,\ref{p3:fig:fit_summary_B311}, but for B164.}
 \end{figure*}

 \begin{figure*}
     \centering
     \includegraphics[width=0.9\textwidth]{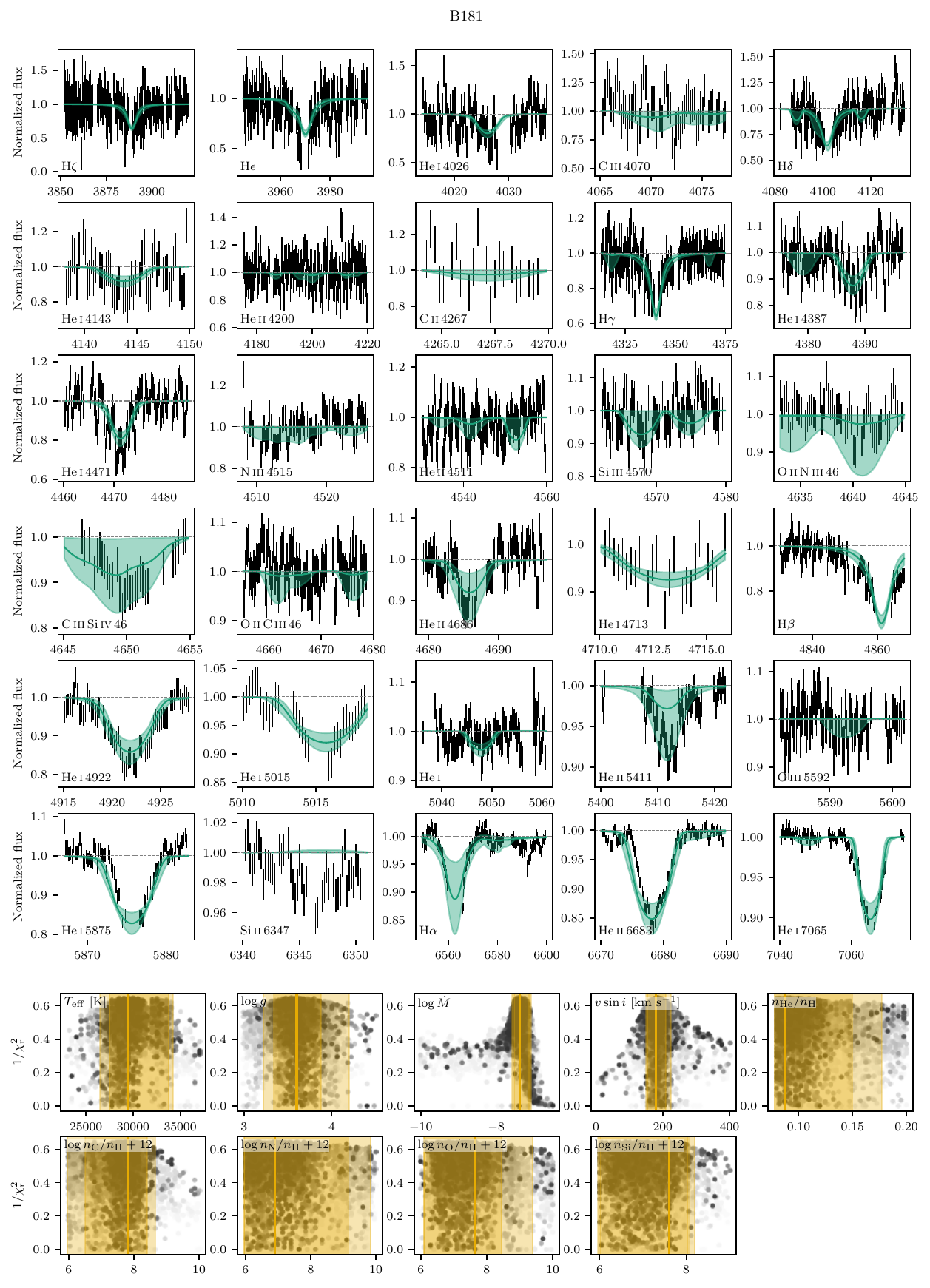}
     \caption{Same as Figure\,\ref{p3:fig:fit_summary_B311}, but for B181.}
 \end{figure*}

 \begin{figure*}
     \centering
     \includegraphics[width=0.9\textwidth]{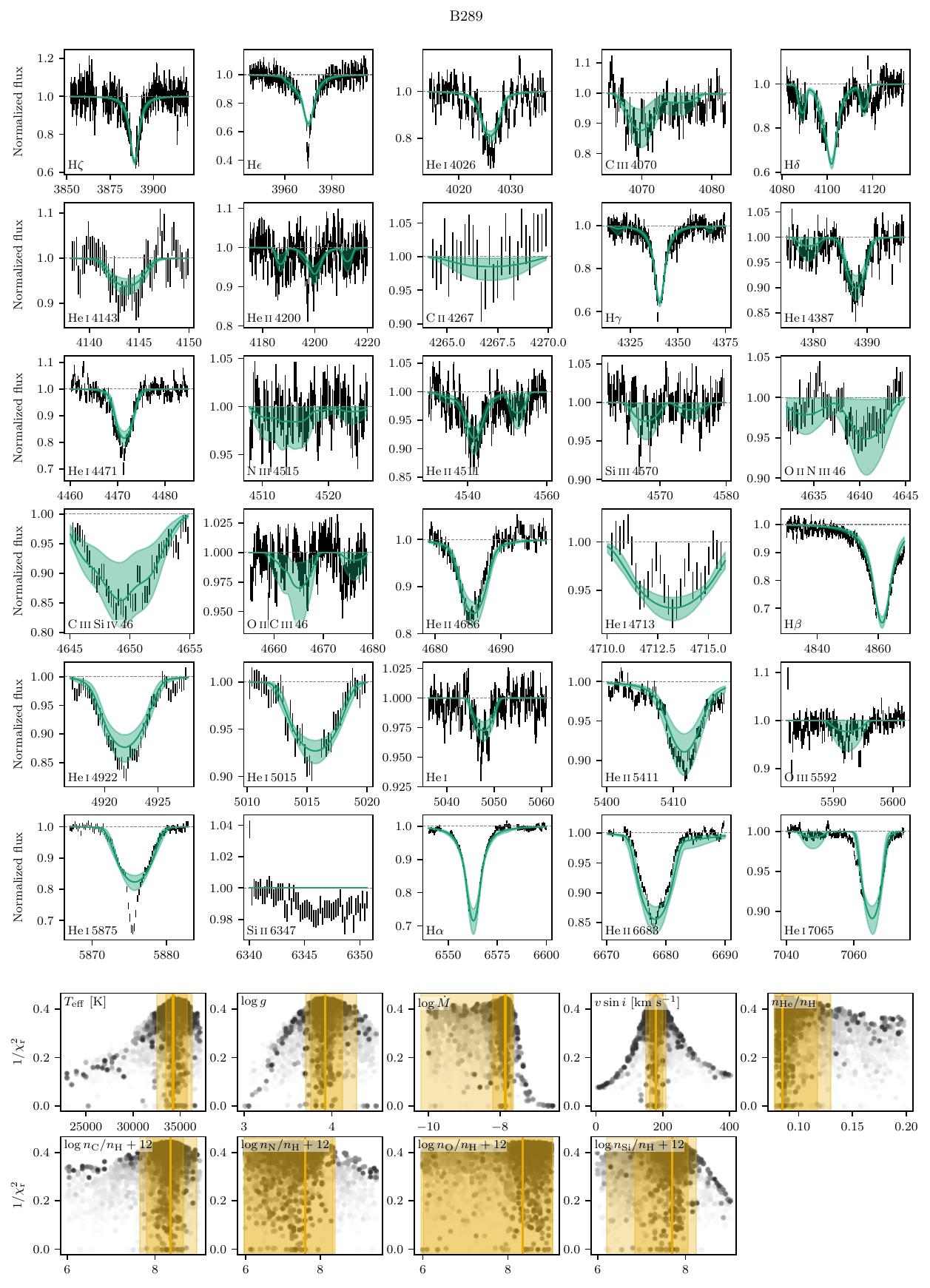}
     \caption{Same as Figure\,\ref{p3:fig:fit_summary_B311}, but for B289.}
 \end{figure*}

\begin{figure*}
     \centering
     \includegraphics[width=0.9\textwidth]{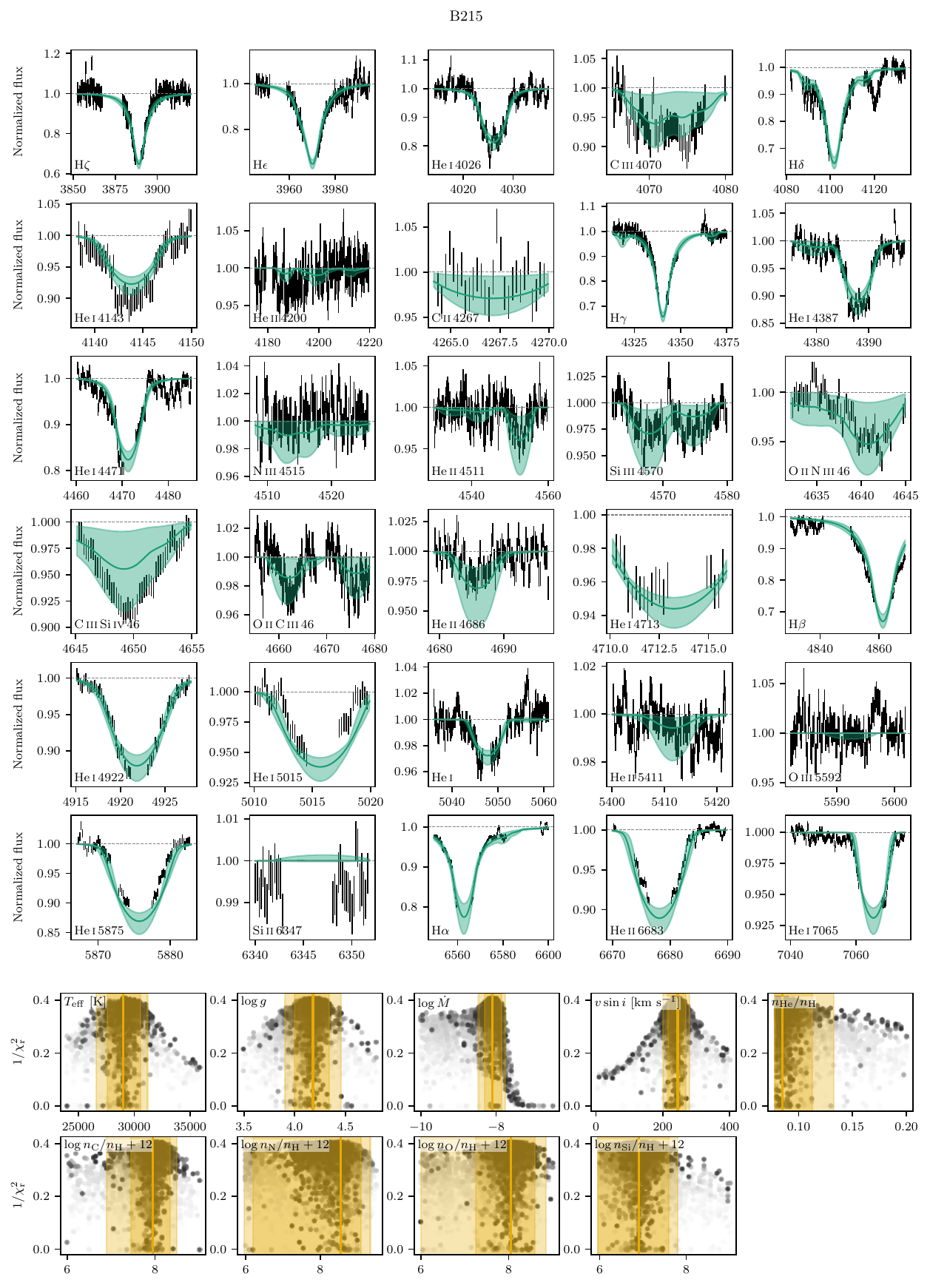}
     \caption{Same as Figure\,\ref{p3:fig:fit_summary_B311}, but for B215.}
 \end{figure*}

\begin{figure*}
     \centering
     \includegraphics[width=0.9\textwidth]{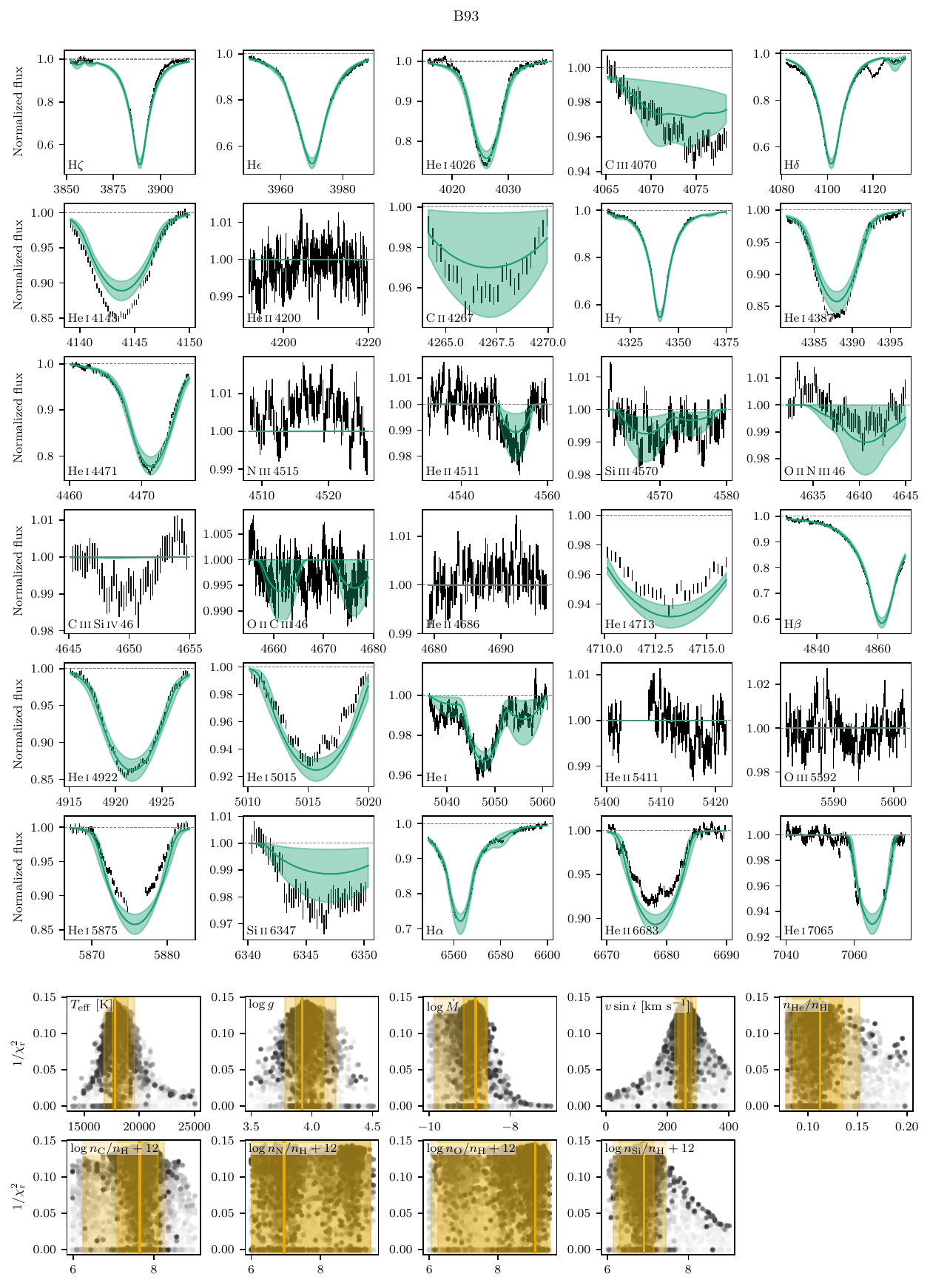}
     \caption{Same as Figure\,\ref{p3:fig:fit_summary_B311}, but for B93.}
 \end{figure*}

\begin{figure*}
     \centering
     \includegraphics[width=0.9\textwidth]{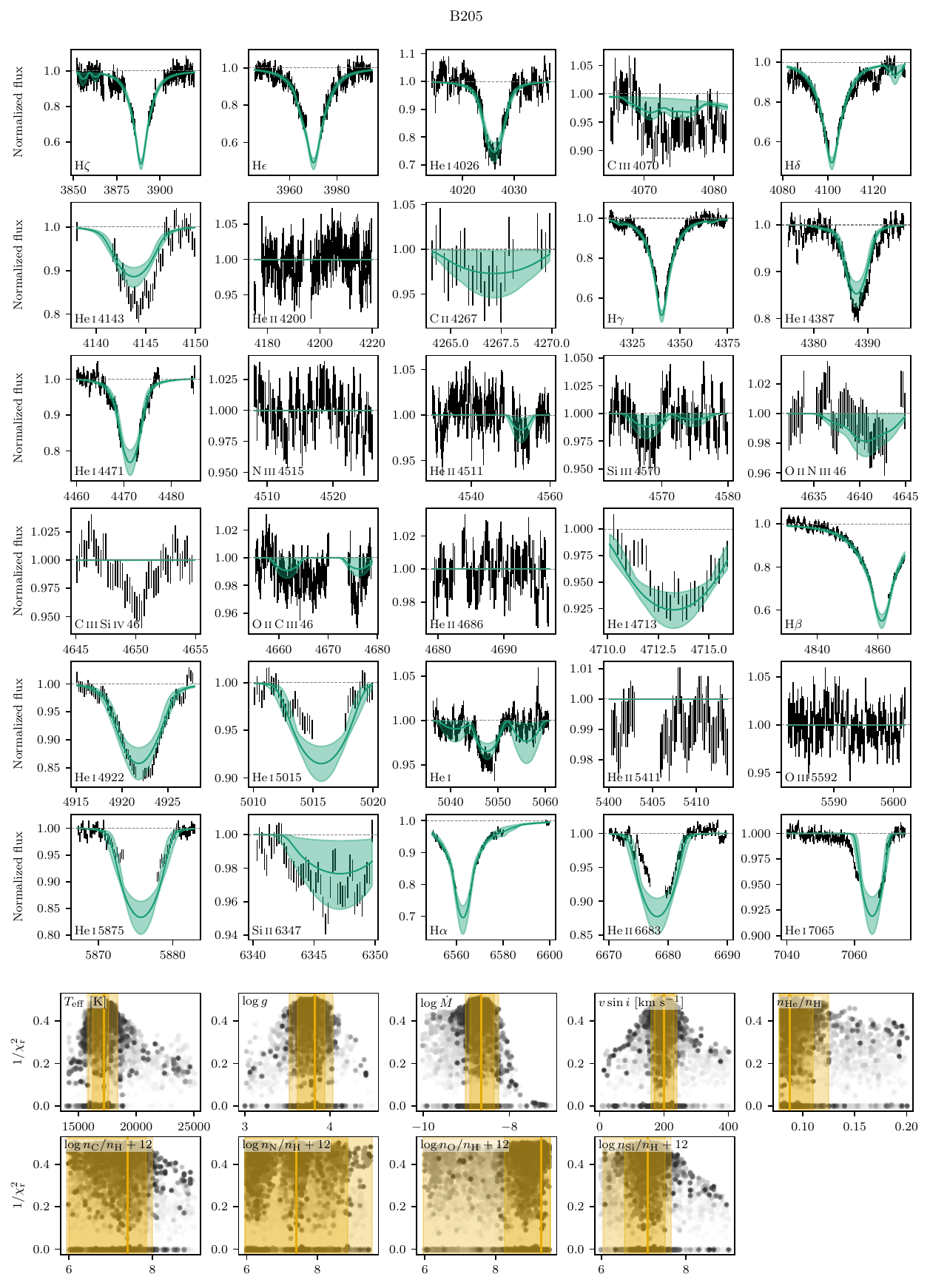}
     \caption{Same as Figure\,\ref{p3:fig:fit_summary_B311}, but for B205.}
 \end{figure*}

\begin{figure*}
     \centering
     \includegraphics[width=0.9\textwidth]{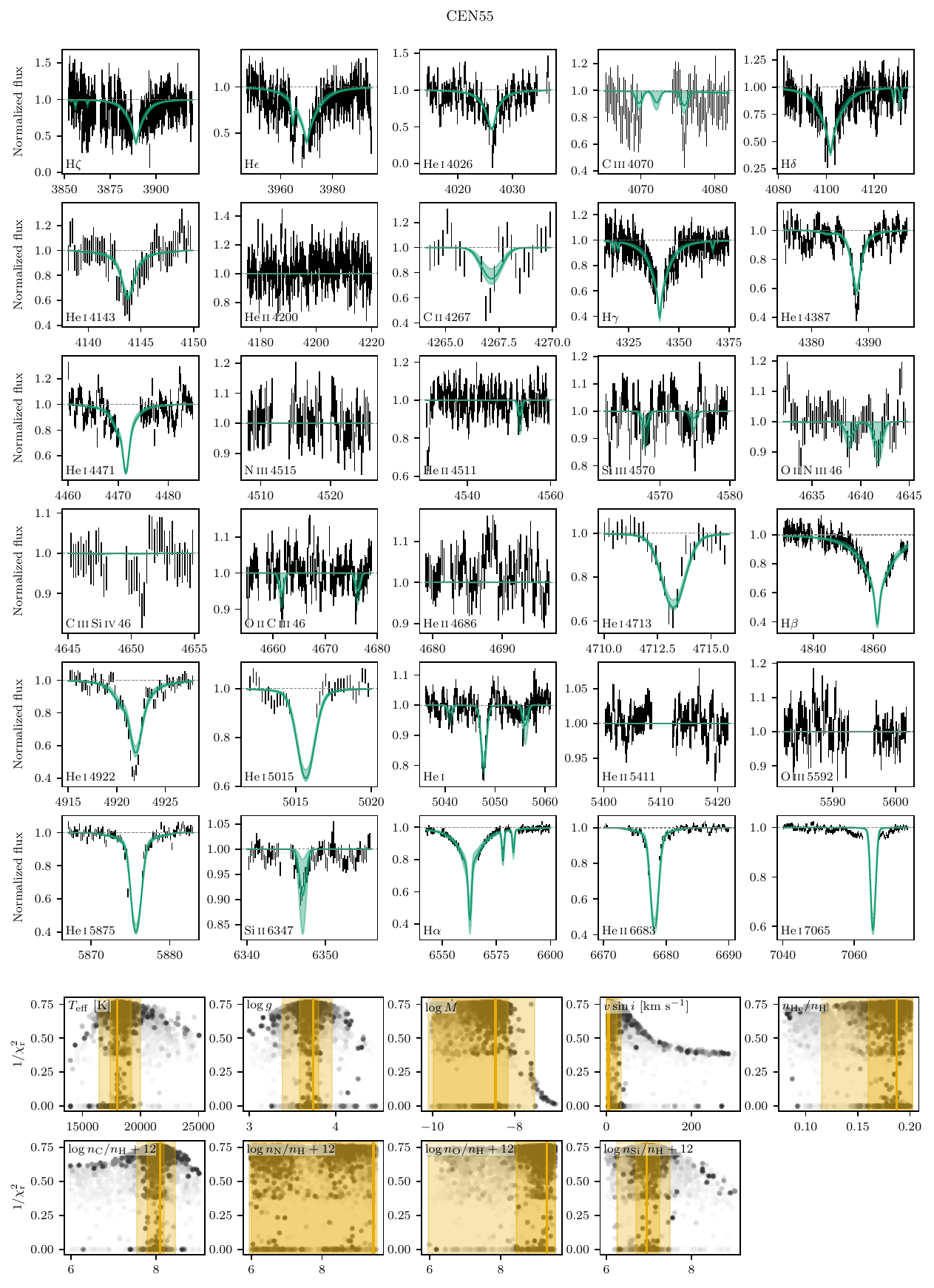}
     \caption{Same as Figure\,\ref{p3:fig:fit_summary_B311}, but for CEN55.}
 \end{figure*}

\begin{figure*}
     \centering
     \includegraphics[width=0.9\textwidth]{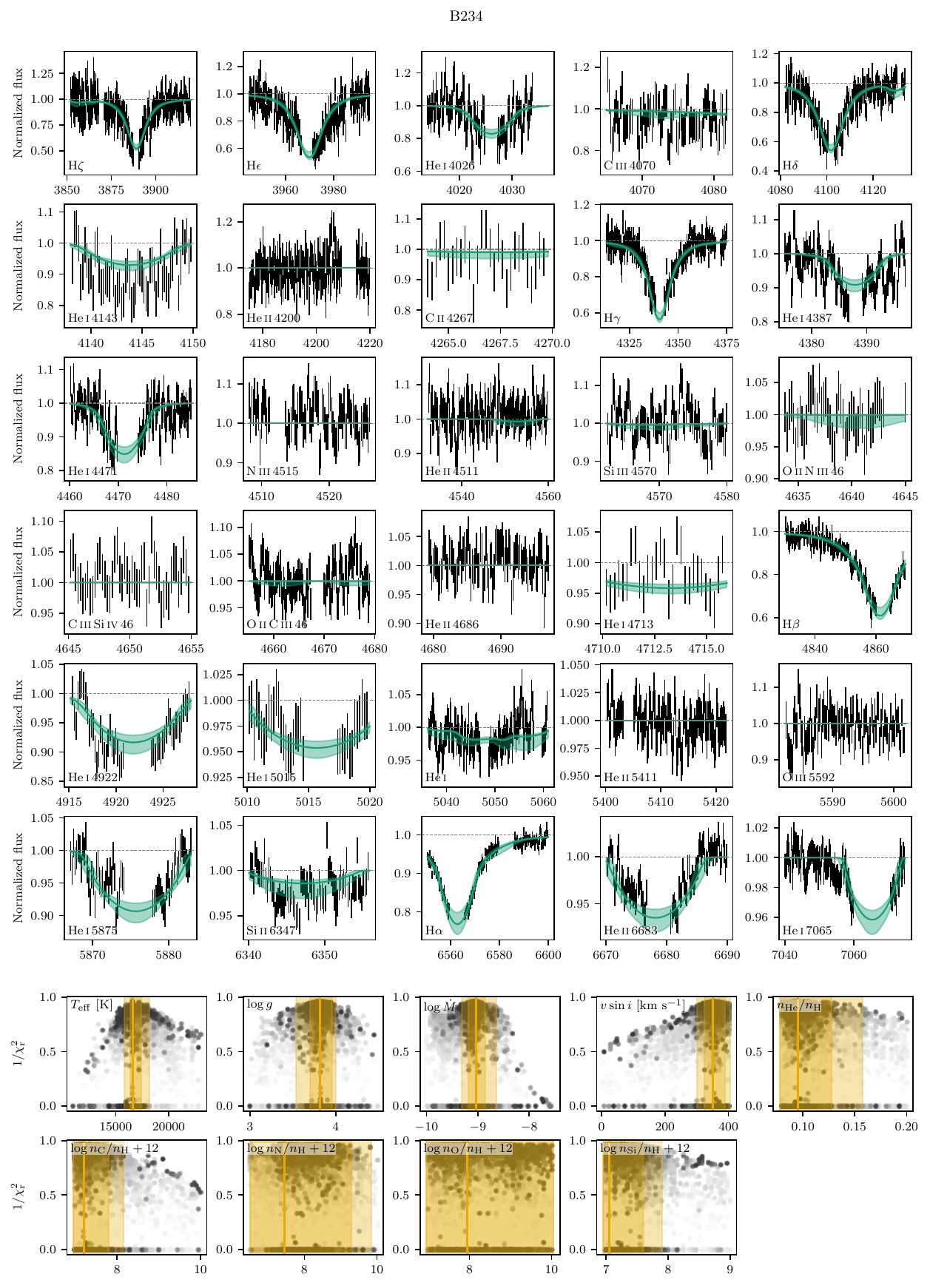}
     \caption{Same as Figure\,\ref{p3:fig:fit_summary_B311}, but for B234.}
 \end{figure*}

\begin{figure*}
     \centering
     \includegraphics[width=0.9\textwidth]{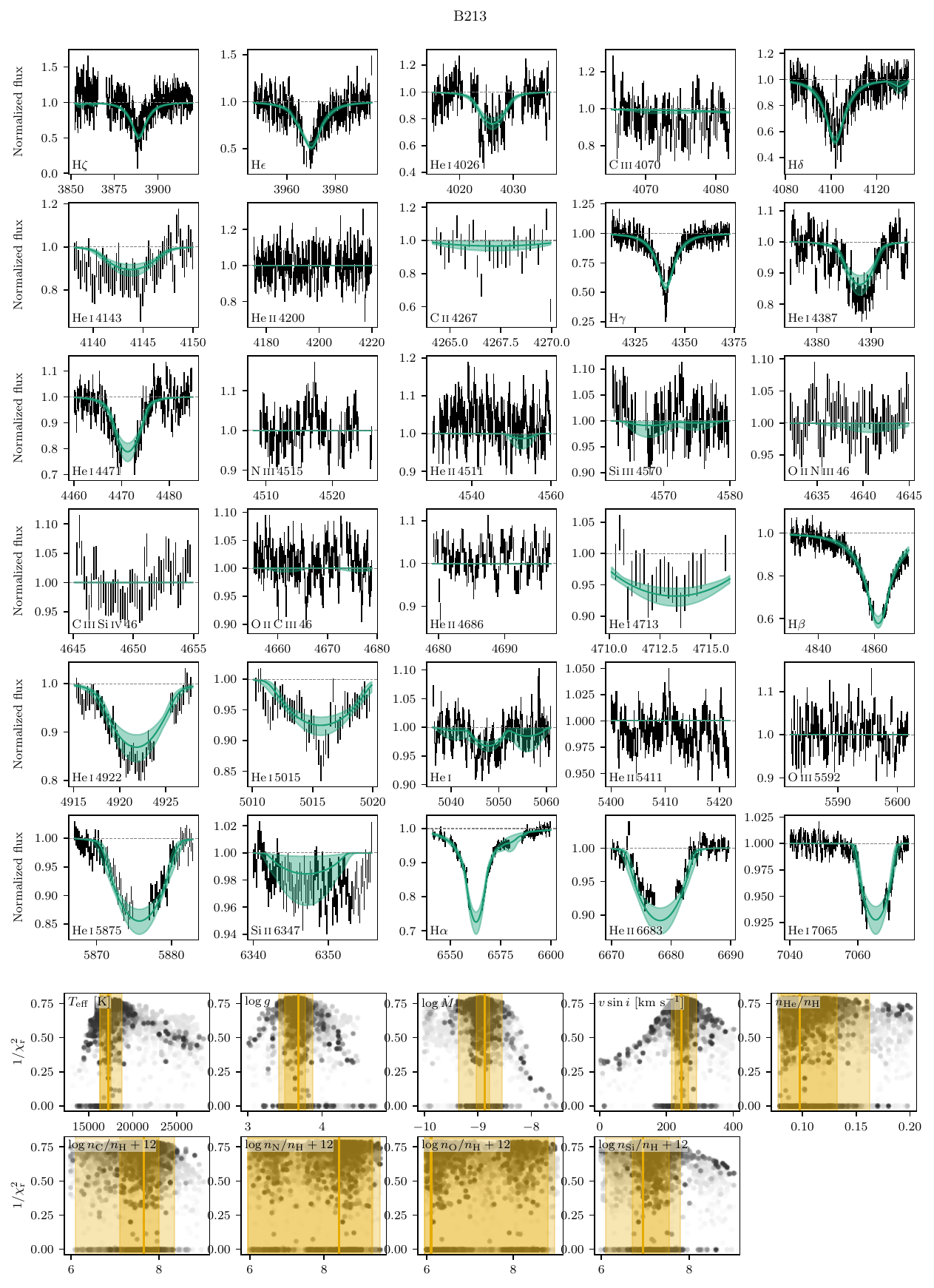}
     \caption{Same as Figure\,\ref{p3:fig:fit_summary_B311}, but for B213.}
 \end{figure*}

\begin{figure*}
     \centering
     \includegraphics[width=\textwidth]{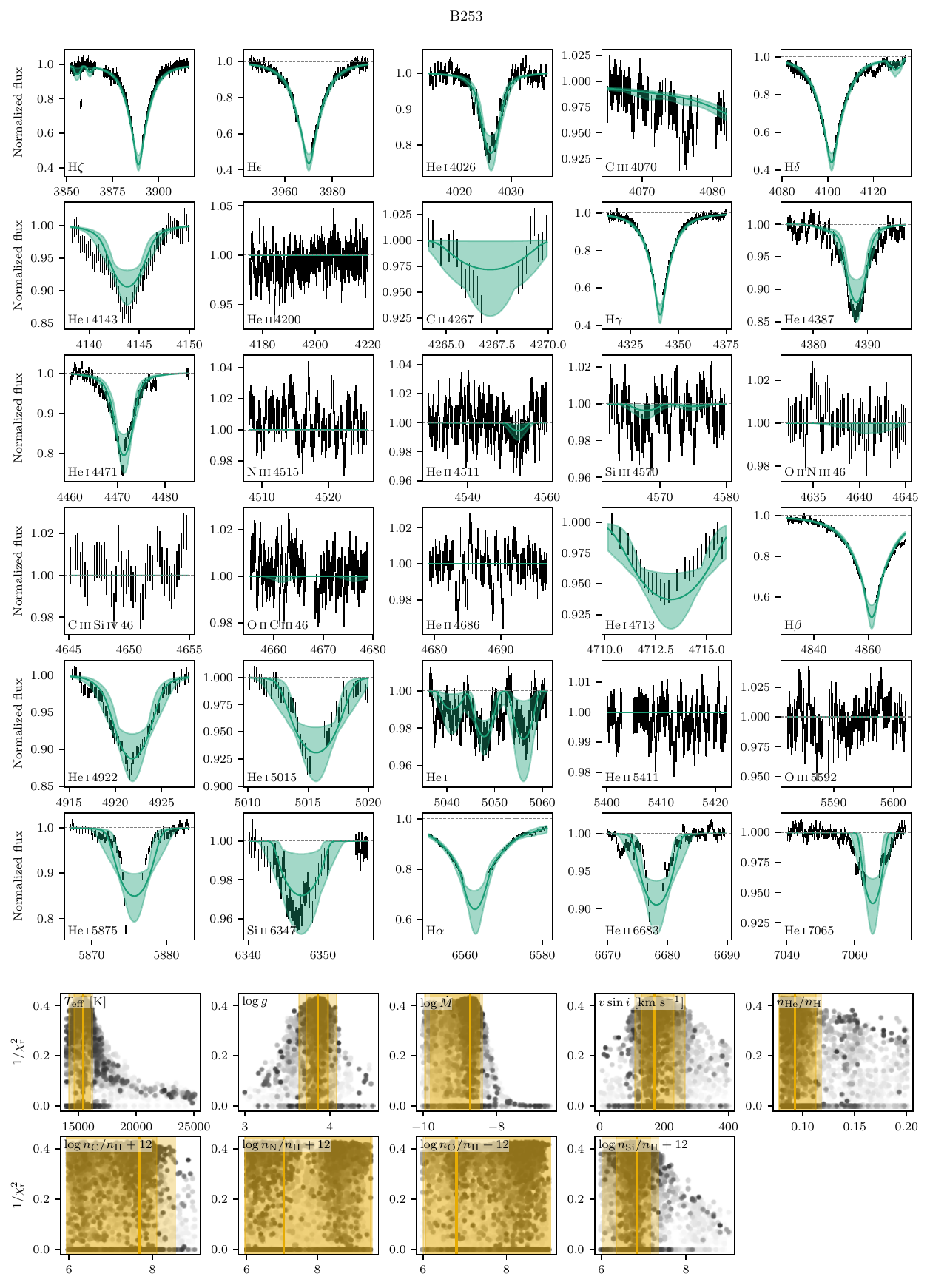}
     \caption{Same as Figure\,\ref{p3:fig:fit_summary_B311}, but for B253.}
 \end{figure*}

\begin{figure*}
     \centering
     \includegraphics[width=0.9\textwidth]{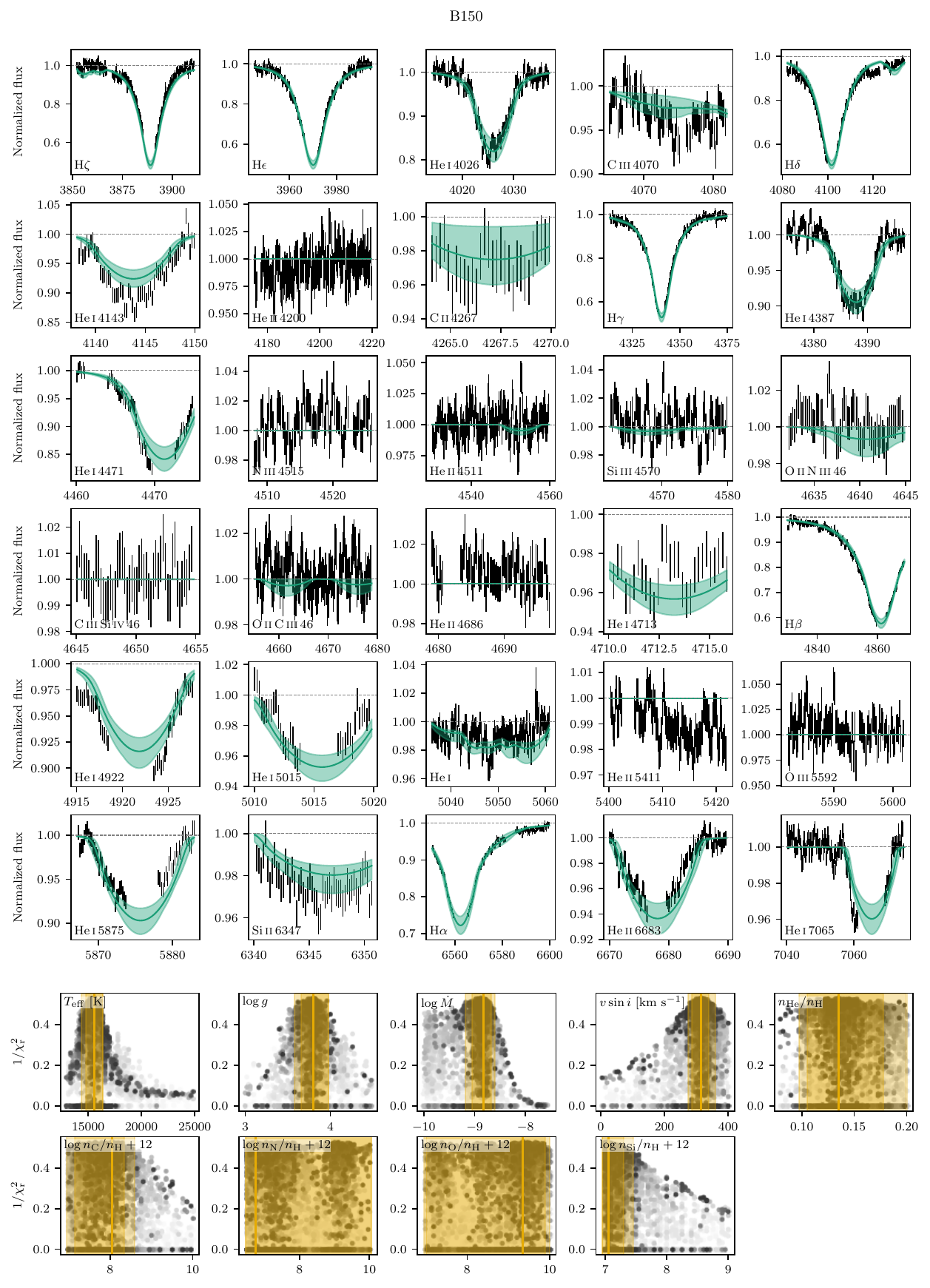}
     \caption{Same as Figure\,\ref{p3:fig:fit_summary_B311}, but for B150.}
 \end{figure*}

\begin{figure*}
     \centering
     \includegraphics[width=0.9\textwidth]{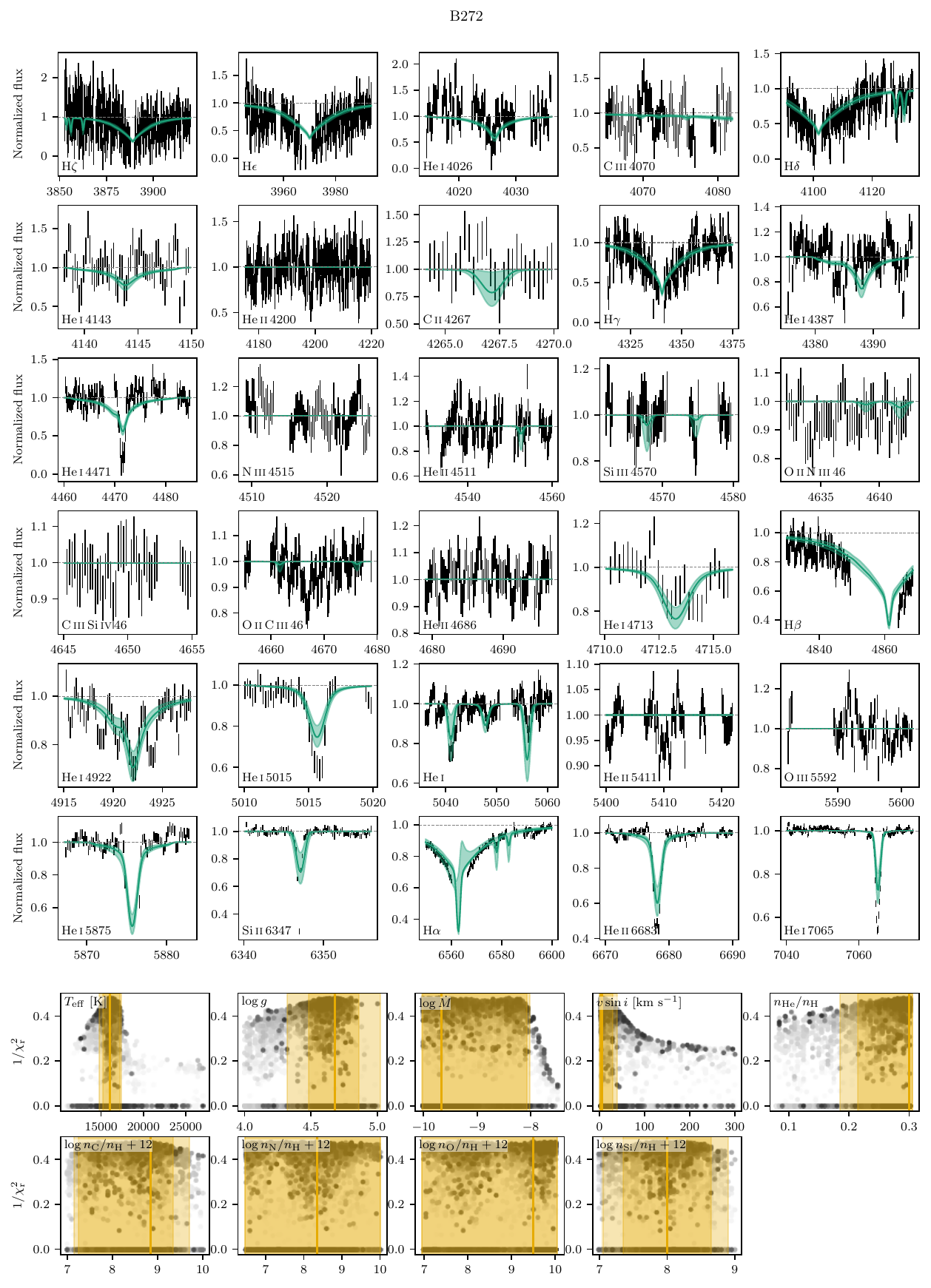}
     \caption{Same as Figure\,\ref{p3:fig:fit_summary_B311}, but for B272.}
 \end{figure*}

\begin{figure*}
     \centering
     \includegraphics[width=0.9\textwidth]{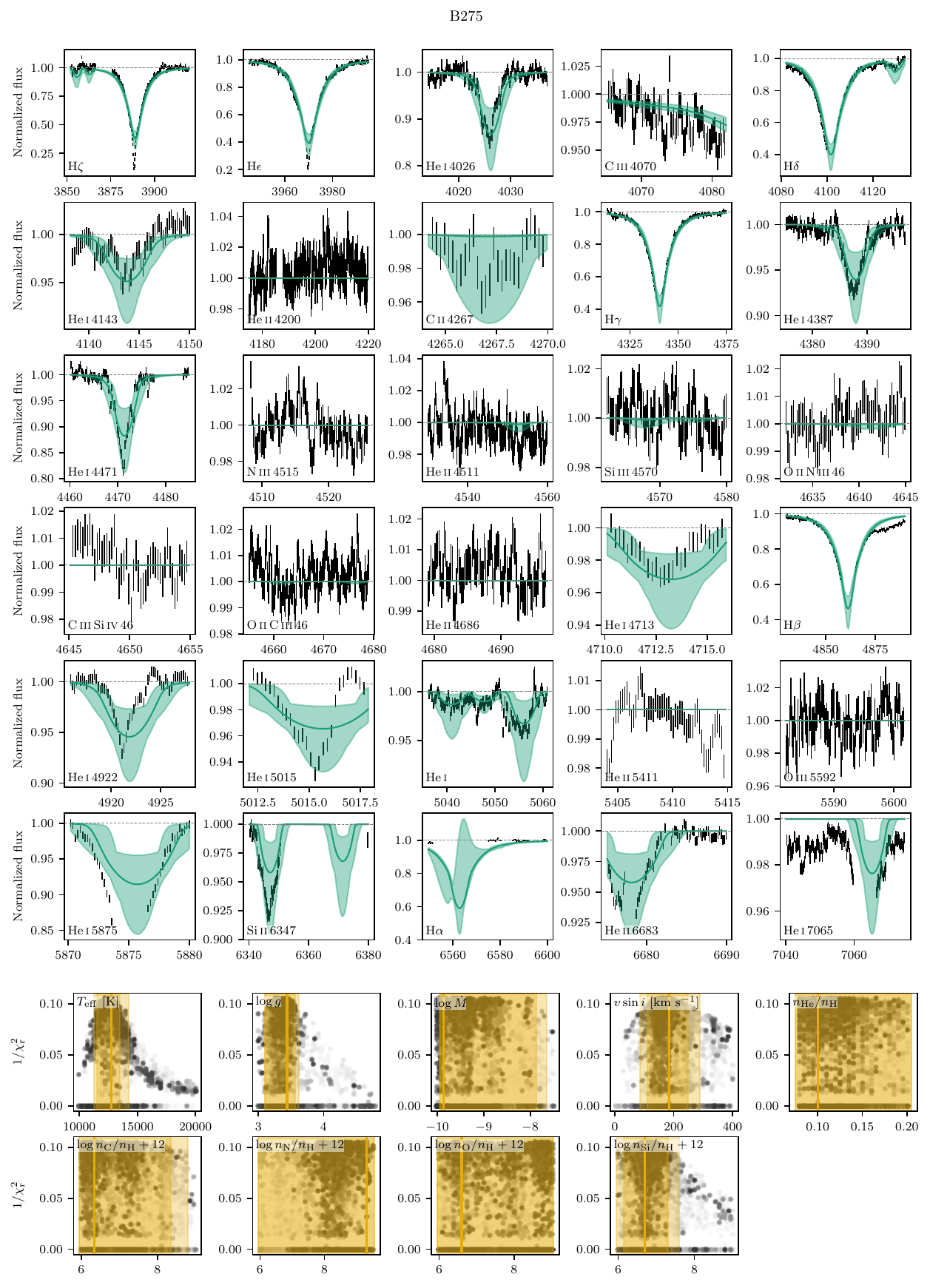}
     \caption{Same as Figure\,\ref{p3:fig:fit_summary_B311}, but for B275.}
 \end{figure*}

\begin{figure*}
     \centering
     \includegraphics[width=0.9\textwidth]{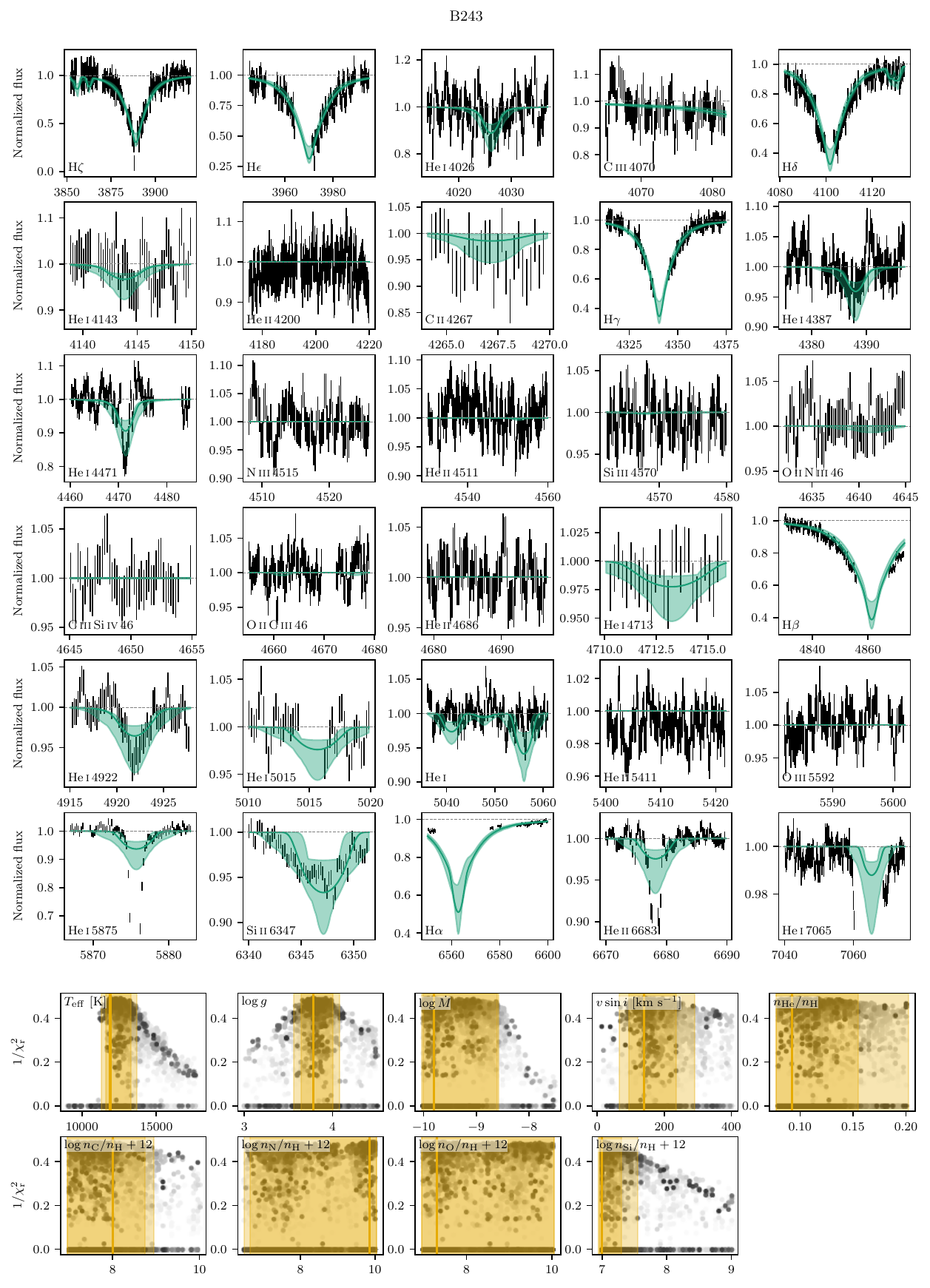}
     \caption{Same as Figure\,\ref{p3:fig:fit_summary_B311}, but for B243.}
 \end{figure*}

\begin{figure*}
     \centering
     \includegraphics[width=0.9\textwidth]{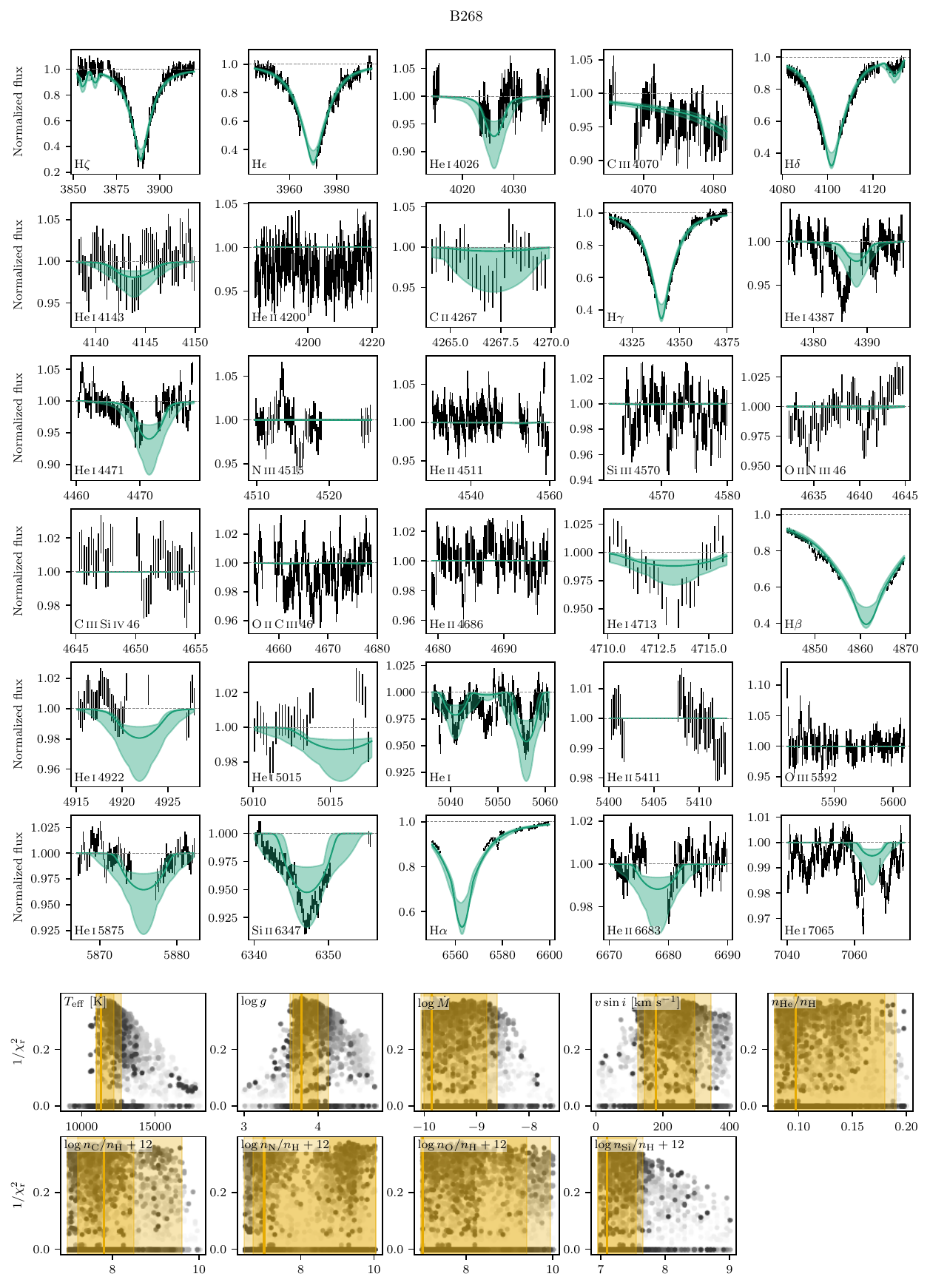}
     \caption{Same as Figure\,\ref{p3:fig:fit_summary_B311}, but for B268.}
 \end{figure*}

\begin{figure*}
     \centering
     \includegraphics[width=0.9\textwidth]{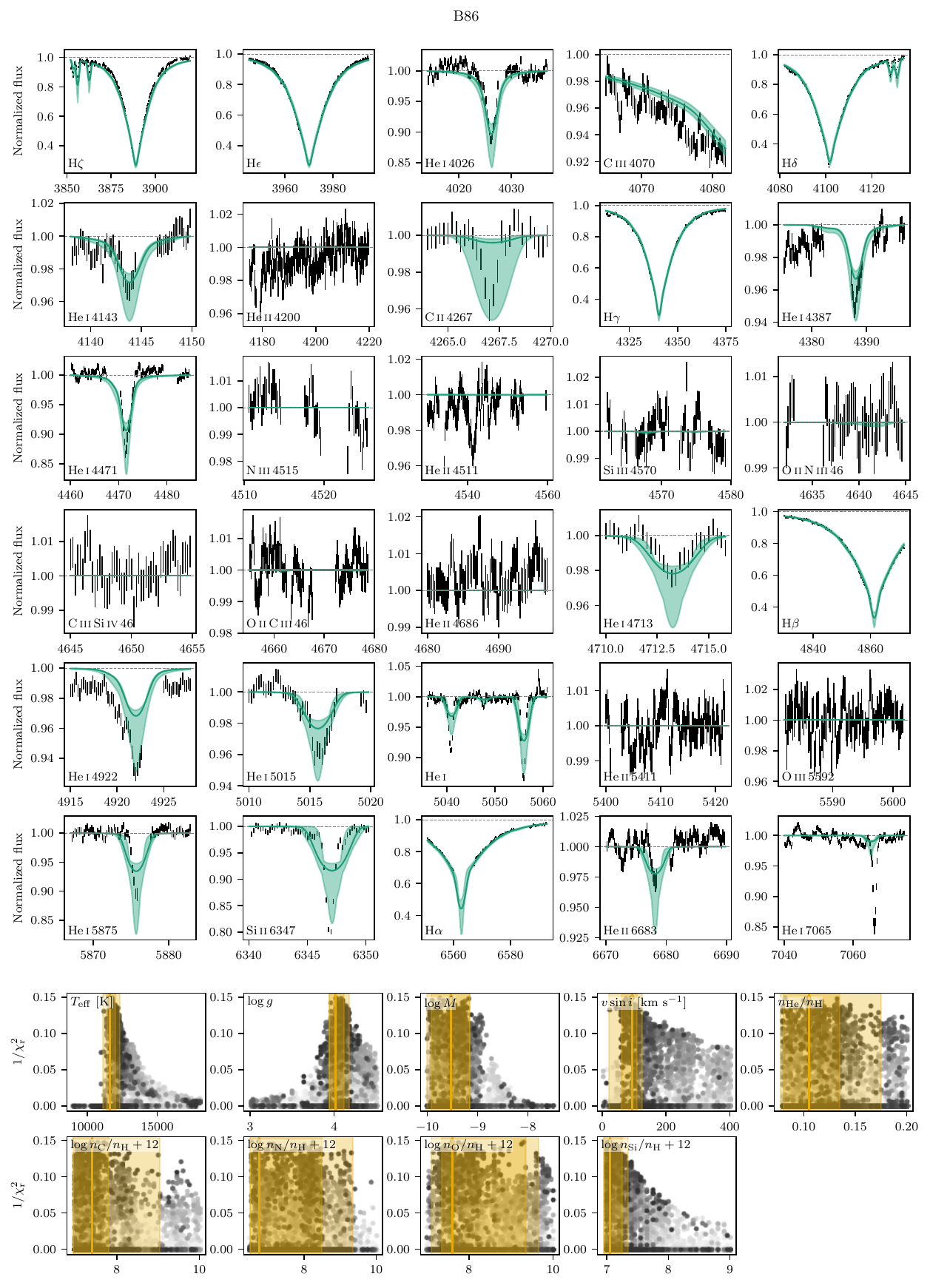}
     \caption{Same as Figure\,\ref{p3:fig:fit_summary_B311}, but for B86.}
 \end{figure*}

\end{appendix}

\end{document}